\documentclass[a4paper,11pt]{article}
\usepackage[margin=2.5cm]{geometry}
\usepackage{amsfonts,amsmath,amssymb}
\usepackage[titletoc,toc,title]{appendix}
\usepackage{authblk}
\usepackage{braket}
\usepackage{cancel}
\usepackage[font={footnotesize,it}]{caption}
\usepackage{cite}
\usepackage{comment}
\usepackage{enumitem}
\usepackage{epsfig}
\usepackage{etoolbox}
\usepackage{float}
\usepackage{graphicx}
\usepackage{epstopdf}
\usepackage{setspace}
\usepackage{hyperref}
\usepackage{cleveref}
\usepackage{orcidlink}
\usepackage{physics}
\usepackage{ragged2e}
\usepackage{soul}
\hypersetup{colorlinks=true,citecolor=red,linkcolor=blue,urlcolor=blue,linktocpage=true}
{\makeatletter\g@addto@macro\bfseries{\boldmath}\makeatother}
\numberwithin{equation}{section}
\crefname{subsection}{subsection}{subsections}
\begin{document}
\title{Covariant odd entanglement entropy in AdS$_3$/CFT$_2$}
\author[1]{Saikat Biswas\,\orcidlink{0009-0001-5545-5200}\thanks{E-mail: {\texttt{saikatb21@iitk.ac.in}}}}
\author[1]{Ankur Dey\,\orcidlink{0009-0001-1077-0442}\thanks{E-mail: {\texttt{ankurd21@iitk.ac.in}}}}
\author[2]{Boudhayan Paul\,\orcidlink{0000-0003-3721-6289}\thanks{E-mail: {\texttt{paul@bimsa.cn}}}}
\author[1]{Gautam Sengupta\,\orcidlink{0000-0002-1118-6926}\thanks{E-mail: {\texttt{sengupta@iitk.ac.in}}}}
\affil[1]{Department of Physics, Indian Institute of Technology, Kanpur 208016, India}
\affil[2]{Beijing Institute of Mathematical Sciences and Applications, Beijing 101408, China}
\date{}
\maketitle
\begin{abstract}
We advance a covariant construction for the holographic odd entanglement entropy (OEE) of time dependent bipartite states in CFT$_2$s dual to bulk AdS$_3$ geometries. In this context we obtain the OEE for bipartite states in zero, finite temperature and finite size CFT$_2$s dual to bulk pure AdS$_3$ and BTZ black hole geometries through appropriate replica techniques. The replica technique results for the time dependent OEE are reproduced modulo constants in the large central charge limit through holographic computations involving the bulk entanglement wedge cross section (EWCS). Subsequently we obtain the time dependent OEE for bipartite states in zero and finite temperature CFT$_2$s with a conserved charge dual to bulk extremal and non-extremal rotating BTZ black holes through both field theory and covariant holographic computations which again match up to constants in the large central charge limit.
\end{abstract}
\clearpage
\setcounter{tocdepth}{2}
\tableofcontents
\clearpage
\section{Introduction}\label{sec_intro}

In recent years quantum entanglement has become an emergent critical research issue across diverse disciplines from many body condensed matter systems to black hole information and quantum gravity. In this context the \textit{entanglement entropy} (EE) characterizes bipartite pure state entanglement although it is inadmissible for mixed states as it involves irrelevant classical and quantum correlations. Several entanglement and correlation measures suitable for bipartite mixed states have been proposed in quantum information theory which include the entanglement negativity \cite{Vidal:2002zz,Plenio:2005cwa}, reflected entropy \cite{Dutta:2019gen,Jeong:2019xdr}, entanglement of purification \cite{Takayanagi:2017knl,Caputa:2018xuf}, and balanced partial entanglement \cite{Wen:2021qgx,Camargo:2022mme,Basu:2022nyl,Wen:2022jxr} amongst others.

Interestingly the authors in \cite{Calabrese:2004eu,Calabrese:2009ez,Calabrese:2009qy} developed a replica technique to compute the EE for bipartite states in two dimensional conformal field theories (CFT$_2$s). In this context Ryu and Takayanagi (RT) advanced a holographic proposal \cite{Ryu:2006bv,Ryu:2006ef} for the EE of a subsystem in a CFT$_d$ dual to bulk AdS$_{d+1}$ geometry through the AdS/CFT framework in terms of the area of the co-dimension two bulk static minimal surface (RT surface) homologous to the subsystem. A covariant generalization of this was subsequently proposed by Hubeny, Rangamani and Takayanagi (HRT) \cite{Hubeny:2007xt} for CFT$_d$s dual to bulk non static AdS$_{d+1}$ geometries involving the area of bulk co-dimension two extremal surfaces. The proofs of these proposals were subsequently described in \cite{Fursaev:2006ih,Headrick:2010zt,Casini:2011kv,Lewkowycz:2013nqa,Dong:2016hjy,Faulkner:2013ana}.

Recently in \cite{Tamaoka:2018ned} another mixed state correlation measure termed \textit{odd entanglement entropy} (OEE) was introduced. The author in \cite{Tamaoka:2018ned} obtained the OEE for bipartite mixed states in a CFT$_2$ through a replica technique and proposed a holographic duality involving the OEE, the EE and the bulk entanglement wedge cross section (EWCS) in the context of the AdS$_3$/CFT$_2$ correspondence. Subsequently the OEE for local and global operator quench states in CFT$_2$s was investigated in \cite{Kusuki:2019rbk,Kusuki:2019evw,BabaeiVelni:2020wfl}. Furthermore in \cite{Kudler-:2020url} the time evolution of the OEE amongst other entanglement and correlation measures and the EWCS were investigated for various quenched CFT$_2$s. Following these, in \cite{Mollabashi:2020ifv} the authors explored the OEE for two-dimensional free scalar field theories. This was extended for $(2+1)$-dimensional Chern-Simons theories and thermofield double (TFD) states in free scalar quantum field theories in \cite{Berthiere:2020ihq} and \cite{Ghasemi:2021jiy} respectively. The OEE for Galilean CFT$_2$s dual to asymptotically flat bulk geometries was obtained in \cite{Basak:2022gcv} in the context of flat holography. Recently the OEE was computed from both field theory and bulk perspectives for $\text{T}\bar{\text{T}}$ deformed holographic CFT$_2$s in \cite{Basu:2023aqz}.

Given the above developments an interesting question in this connection is to address the significant issue of time dependent OEE for bipartite states in CFT$_2$s and its covariant holographic description involving the extremal EWCS
from the bulk dual non static AdS$_3$ geometries. Such an exercise for the time dependence and evolution of the OEE is expected to be of critical relevance in addressing mixed state entanglement structures in quantum field theories and to the recent progress towards a resolution of the black hole information loss paradox in the framework of the quantum extremal islands. Furthermore this would also provide crucial insights towards a derivation of the holographic duality involving the OEE, EE and the EWCS mentioned above, from a bulk gravitational path integral along the lines of \cite{Dong:2021clv}, which is an outstanding open issue. In this article we address this crucial issue and first develop a comprehensive replica construction for the OEE of various time dependent bipartite states of two disjoint intervals, two adjacent intervals, and a single interval in zero and finite temperature and finite size holographic CFT$_2$s. In this context we obtain the OEE for such time dependent bipartite states in CFT$_2$s dual to pure AdS$_3$ geometries (in Poincar{\'e} and global coordinates) and BTZ black holes respectively. The field theory replica technique results in the large central charge limit are then verified modulo certain constants through the holographic duality proposed in \cite{Tamaoka:2018ned} involving the covariant construction of the extremal EWCS for the corresponding dual bulk geometries. Furthermore we obtain the OEE for bipartite states in zero and finite temperature CFT$_2$s with a conserved charge dual to bulk extremal and non-extremal rotating BTZ black holes respectively, through both field theory and holographic computations involving the extremal EWCS. The replica technique results in the large central charge limit again match with the corresponding covariant holographic computations modulo constants.\footnote {We would like to emphasize here that our covariant construction for the time dependent OEE described in this article are with reference to usual holographic CFT$_2$s in contrast to  \cite{Kudler-:2020url} which dealt with quenched systems involving boundary CFT$_2$s (BCFT$_2$s).}

This article is organized as follows. In \cref{sec_review} we briefly review the basic features of the OEE and the covariant HRT proposal. In \cref{sec_nr_cft} we obtain the time dependent OEE in CFT$_2$s dual to pure AdS$_3$ spacetimes and BTZ black holes through a replica technique. These results are then verified through the corresponding covariant holographic computations in \cref{sec_nr_holo}. Following a similar approach, we compute the time dependent OEE in CFT$_2$s dual to bulk non-extremal and extremal rotating BTZ black holes in \cref{CFT_OEE_R_SYS}, while the corresponding covariant holographic description is described in \cref{HOL_OEE_R_SYS}. Finally in \cref{SUMMaRY} we summarize our results and present our conclusions.

\section{Review of earlier literature}\label{sec_review}
	
\subsection{Odd entanglement entropy}

In this subsection we briefly review the definition of the odd entanglement entropy (OEE) for bipartite mixed states introduced in \cite{Tamaoka:2018ned}. In this context a bipartite mixed state configuration $A \cup B$ of subsystems $A$ and $B$ on Hilbert spaces $\mathcal{H}_{A}$ and $\mathcal{H}_{B}$ respectively is considered, where the reduced density matrix\footnote{Note that in this case $(A \cup B)^c$ has been traced over.} $\rho_{A B}$ is defined on the Hilbert space $\mathcal{H}_{AB}=\mathcal{H}_{A} \otimes \mathcal{H}_{B}$. The partial transposition of the reduced density matrix described as $\rho^{T_B}_{A B}$ with respect to the subsystem $B$ is then defined as 
	\begin{align} \label{eq_partialtranspose}
		\left< e_i^{(A)} e_j^{(B)} \Big| \rho_{A B}^{T_{B}} \Big| e_k^{(A)} e_l^{(B)} \right> = \left< e_i^{(A)} e_l^{(B)} \Big| \rho_{A B} \Big| e_k^{(A)} e_j^{(B)} \right>,
	\end{align}
where $\big|e_i^{(A)}\big>$ and $\big|e_j^{(B)}\big>$ signify the bases for the Hilbert spaces $\mathcal{H}_{A}$ and $\mathcal{H}_{B}$ respectively.
The \textit{R\'enyi odd entanglement entropy} of order $ n_o $ (where $ n_o $ is an odd positive integer) has the following definition \cite{Basak:2022gcv}
	\begin{align} \label{eq_Renyi_oee}
		S_o^{(n_o)}(A:B)=\frac{1}{1-n_o} \log \left[ \text{Tr} \left( \rho_{AB}^{T_{B}} \right)^{n_o} \right].
	\end{align}
The OEE may then be obtained in terms of the R\'enyi odd entanglement entropy through the analytic continuation $n_o \to 1$ as follows \cite{Tamaoka:2018ned}
	\begin{align}\label{eq_vn_oee}
		S_o(A:B)=\left.\lim_{n_o \to 1} \middle[S_o^{(n_o)}(A:B)\right].
	\end{align}	
	
Several quantum information theoretic properties of the OEE  have been mentioned in the literature and demonstrated numerically for specific examples in \mbox{\cite{Tamaoka:2018ned,Basak:2022gcv,Mollabashi:2020ifv}}, however analytic derivation of these are involved and an outstanding issue for future investigations. These properties are as follows
	
\begin{itemize}
\item $S_o (A:B)\geq 0$ (positive semi-definiteness),
\item $S_o (A:B_1 B_2)\geq S_o (A:B_1)$ (monotonicity),
\item $S_o (A:B_1 B_2)\leq S_o (A:B_1)+S_o (A:B_2)$ (polygamy relation),
\item $S_o (A_1 A_2:B_1 B_2)\geq S_o (A_1:B_1)+S_o (A_2:B_2)$ (breaking of strong super additivity).
\end{itemize}

The above properties are expected to be valid for the OEE  to serve as a bipartite mixed state correlation measure in quantum information theory.

\subsection{OEE in a CFT$_2$}
	
We now describe the replica technique utilized in \cite{Tamaoka:2018ned} to compute the OEE for bipartite states in CFT$_2$s. In this context the trace $\text{Tr} \left( \rho_{AB}^{T_{B}} \right)^{n_o}$ in  \cref{eq_Renyi_oee} may be expressed as a four-point correlator of twist fields on the complex plane as follows \cite{Tamaoka:2018ned, Calabrese:2012nk}
	\begin{align}\label{eq_4pf}
		\text{Tr} \left( \rho_{AB}^{T_{B}} \right)^{n_o} = \left<\sigma_{n_o}(z_1,\bar{z}_1)\bar{\sigma}_{n_o}(z_2,\bar{z}_2)\bar{\sigma}_{n_o}(z_3,\bar{z}_3)\sigma_{n_o}(z_4,\bar{z}_4)\right>,
	\end{align}
where $[z_1,z_2]$ and $[z_3,z_4]$  describe two disjoint boosted spatial intervals specifying the subsystems $A$ and $B$. Here $\sigma_{n_o}$ and $\bar{\sigma}_{n_o}$ represent the twist and anti-twist operators in the CFT$_2$ respectively with conformal dimensions \cite{Calabrese:2012ew,Calabrese:2012nk}
	\begin{align}\label{eq_hcdimension}
		h_{n_o}=\bar{h}_{n_o}=\frac{c}{24} \left( n_o-\frac{1}{n_o} \right),
	\end{align}
where $c$ is the central charge. For later computations we note here that the conformal dimensions of the twist operators $\sigma^2_{n_o}$ and $\bar{\sigma}^2_{n_o}$ are given as \cite{Calabrese:2012ew,Calabrese:2012nk,Tamaoka:2018ned}
	\begin{align}\label{eq_h2cdimension}
		h^{(2)}_{n_o}=\bar{h}^{(2)}_{n_o}=\frac{c}{24} \left( n_o-\frac{1}{n_o} \right).
	\end{align}

\subsection{Holographic OEE in a CFT$_2$}

We begin this subsection with brief discussions of the RT and HRT proposals for the holographic entanglement entropy of subregions in the dual CFT. According to the RT conjecture \cite{Ryu:2006bv,Ryu:2006ef}, the holographic entanglement entropy (HEE) of a spatial subregion $A$ in a CFT$_d$ dual to a bulk static AdS$_{d+1}$ spacetime is described by
	\begin{align}
	S_{A}=\frac{\text{Area}\,\left(\Sigma_{A}^{min}\right)}{4G_N^{(d+1)}},
	\end{align}
where $\Sigma_{A}^{min}$ represents the minimal surface (RT surface) homologous to $A$, and $G_N^{(d+1)}$ is the Newton constant in $(d+1)$ dimensions. For the AdS$_3$/CFT$_2$ scenario, the RT surfaces are given by geodesics homologous to the corresponding intervals in the dual CFT.

The covariant version of the HEE is described by the HRT conjecture \cite{Hubeny:2007xt} which states that the HEE of a time dependent subregion $A_t$ in a CFT$_d$ dual to a bulk non static AdS$_{d+1}$ geometry at a time $t$ is given as
	\begin{align}\label{HRT}
	S_{A_t}=\frac{\text{Area}\,\left(\mathcal{Y}_{A_t}^{ext}\right)}{4G_N^{(d+1)}},
	\end{align}
where $\mathcal{Y}_{A_t}^{ext}$ is the extremal surface (HRT surface) homologous to $A_t$, characterized through the Bousso light-sheet construction \cite{Bousso:1999xy,Bousso:1999cb,Bousso:2002hon}. For the AdS$_3$/CFT$_2$ framework, the HRT surfaces reduce to extremal curves (geodesics) homologous to the corresponding intervals on the boundary.

In the AdS$_3$/CFT$_2$ context, a holographic duality characterizing the OEE, HEE and the bulk EWCS of the bipartite system $A\cup B$ was proposed in \cite{Tamaoka:2018ned} as follows
	\begin{align}\label{eq_holodual}
	S_o(A:B)=E_W(A:B)+S(A \cup B),
	\end{align}
where $S(A \cup B)$ denotes the HEE for the subsystem ($A \cup B$), and $E_W(A:B)$ denotes the bulk EWCS for the subregion ($A \cup B$). The relation in \cref{eq_holodual} may be regarded as the definition of the holographic OEE between the subsystem $A$ and $B$. 
 
It is important to note however that the proper minimization procedure to obtain the holographic OEE is expected to involve the minimization of the sum of two different surfaces, namely the RT surface homologous to the boundary subsystem (\mbox{$A \cup B$}) and the surface dividing the entanglement wedge of the two subsystems. However, in \mbox{\cref{eq_holodual}} the holographic OEE is presented as the sum of two minimized quantities namely the EWCS and the HEE. As described in \mbox{\cite{Kusuki:2019rbk,Kusuki:2019evw}}, these two minimization procedures coincide in the large \mbox{$c$} limit as they correspond to the evaluation of the same dominant single conformal block in the dual CFT$_2$. Away from this limit, the two distinct minimization procedures described above may lead to different results as the corresponding dominant single conformal block may be different.
	
In the present article we extend the definition of the holographic OEE in \mbox{\cref{eq_holodual}} to time dependent scenarios through the covariant HRT prescription in \mbox{\cref{HRT}} and the definition of the extremal EWCS \mbox{\cite {Kusuki:2019zsp,Kudler-Flam:2018qjo,KumarBasak:2020eia}}. For the \mbox{AdS$_3$/CFT$_2$} scenarios considered here, the two different extremization procedures mentioned above should also converge for time dependent scenarios in the large \mbox{$c$} holographic limit as the dominant single conformal blocks corresponding to these are expected to be identical as earlier. This is because the field theory replica technique computations and the large $c$ approximations are formally similar for both static and time dependent scenarios. However an analytic proof for this assertion is still an open issue.
	
We should also mention here that though \mbox{\cref{eq_holodual}} implies that the difference between OEE and EE leads to the EWCS which is half of the reflected entropy between the subsystems \mbox{$A$} and \mbox{$B$} in the holographic limit, they are distinct measures in quantum information theory. Their equivalence in the holographic limit follows from the universality of the semi-classical regime probed in holography and as a consequence the bulk EWCS has been proposed as the possible dual of several entanglement and correlation measures such as the entanglement negativity, reflected entropy and the entanglement of purification, although they are distinguishable measures in quantum information theory and are expected to be different for general quantum field theories.
	
The above issue has been investigated in the context of the OEE in\mbox{\cite{Mollabashi:2020ifv}} through the investigation of the nature of correlations involved in the difference \mbox{$S_o(A: B)-S (A \cup B)$} of bipartite mixed states for quantum field theories in various regimes. It was observed for specific examples that the above difference increases for regimes with large quantum correlations but decreases with the increase of classical correlations consistent with holographic field theories. However for regimes with large classical correlations this quantity was rendered insensitive also to classical correlations. In other cases for certain regimes the difference \mbox{$S_o(A: B)-S (A \cup B)$} reduced to the logarithmic negativity, which describes the upper bound on the distillable entanglement of a bipartite mixed state, illustrating that it was dominated by quantum correlations. It appears that the difference \mbox{$S_o (A:B)-S (A \cup B)$} measures the total correlations for a mixed state including both classical and quantum correlations, however depending on the regimes of the parameter space of the field theories one or the other dominates. This is an extremely interesting issue which requires further investigations in the context of general quantum field theories. Having briefly recapitulated the essential elements required for our analysis, in the next section we describe our investigation of the time dependent OEE for bipartite states in CFT$_2$s.

	\section{Time dependent OEE in CFT$_2$s}\label{sec_nr_cft}

	\subsection{OEE for two disjoint intervals}\label{ssec_nr_cft_dis}
	
	In this subsection we compute the time dependent OEE for the bipartite mixed state configurations of two boosted disjoint intervals $A_1$ and $A_2$ in CFT$_2$s at zero and finite temperatures and for finite size through a replica technique.

	\subsubsection{Two disjoint intervals at zero temperature}\label{sssec_nr_cft_dis_zt}

	We first consider a mixed state configuration of two boosted disjoint intervals $A_1=[z_1,z_2]$ and $A_2=[z_3,z_4]$ at zero temperature, where $z_i=x_i+i\tau_i$. The relevant twist correlator for the OEE of this configuration is a four-point function, which may be expressed in terms of the $t$-channel conformal blocks as \cite{Tamaoka:2018ned,Fitzpatrick:2014vua}
	\begin{align}\label{eq_4pf_dis}
		\left<\sigma_{n_o}(z_1,\bar{z}_1)\bar{\sigma}_{n_o}(z_2,\bar{z}_2)\bar{\sigma}_{n_o}(z_3,\bar{z}_3)\sigma_{n_o}(z_4,\bar{z}_4)\right> & =  \sum_p b_p \,   \mathcal{F}(c,h_{n_o},h_p,1-\eta) \, \mathcal{\bar{F}}(c,\bar{h}_{n_o},\bar{h}_p,1-\bar{\eta}) \notag \\ &  \times \left((z_4-z_1)(z_3-z_2)\right)^{-2h_{n_o}} \left((\bar{z}_4-\bar{z}_1)(\bar{z}_3-\bar{z}_2)\right)^{-2\bar{h}_{n_o}}.
	\end{align}
	The twist operator $\sigma^{(2)}_{n_o}$ of conformal weight $h^{(2)}_{n_o}=h_{n_o}$ has the most dominant contribution in the $t$-channel \cite{Fitzpatrick:2014vua,Tamaoka:2018ned,Basak:2022gcv}. The conformal block for $\sigma^{(2)}_{n_o}$ in this case may be expressed in the large central charge limit as	
	\begin{align}\label{eq_4pf_dis_cb}
		\log \mathcal{F} (c,h_{n_o}, h_{n_o}^{(2)}, 1-\eta) = -h_{n_o}^{(2)}\left( \log \left[ \frac{1+ \sqrt{\eta}}{1- \sqrt{\eta}} \right]-\log 2\right),
	\end{align}
	where $\eta=\frac{(z_2-z_1)(z_4-z_3)}{(z_3-z_1)(z_4-z_2)}$ and $\bar{\eta}=\frac{(\bar{z}_2-\bar{z}_1)(\bar{z}_4-\bar{z}_3)}{(\bar{z}_3-\bar{z}_1)(\bar{z}_4-\bar{z}_2)}$ are the cross ratios. Using \cref{eq_Renyi_oee,eq_vn_oee,eq_4pf,eq_4pf_dis,eq_4pf_dis_cb}, the OEE for the mixed state configuration of two disjoint intervals at zero temperature for a large central charge $c$ may be obtained as
	\begin{align}\label{eq_nr_cft_dis_zt_oee}
		S_o(A_1:A_2) & =   \frac{c}{12}  \log \left[ \frac{1+ \sqrt{\eta}}{1- \sqrt{\eta}} \right]   +\frac{c}{12}\log \left[ \frac{1+ \sqrt{\bar{\eta}}}{1- \sqrt{\bar{\eta}}} \right] +\frac{c}{3} \log \left[ \frac {\sqrt{x_{41}^2-t_{41}^2} }{a} \right] \notag \\ &  +\frac{c}{3} \log \left[ \frac{ \sqrt{x_{32}^2-t_{32}^2} }{a} \right]-\frac{c}{6} \log 2.
	\end{align}
	Here $x_{ji} = (x_j-x_i)$ and $t_{ji} = (t_j-t_i)$, where the Euclidean time $\tau_j$ has been replaced with $\tau_j = -it_j$, and $a$ is the UV cut-off. Note that the constant term arises from the conformal block expansion and there may be other undetermined OPE constants in the above equation.

	\subsubsection{Two disjoint intervals in a finite size system}\label{sssec_nr_cft_dis_fs}

	In this case the two boosted disjoint intervals are defined in a CFT$_2$ described on a cylinder of circumference $L$ along the spatial direction. This may be obtained through a conformal map from the complex plane (described by the coordinate $z$) to a cylinder (described by $\omega$) as follows \cite{Calabrese:2009qy,Calabrese:2012nk}
\begin{align}\label{eq_nr_fstrans}
		z \rightarrow \omega=e^{\frac{-i2\pi z}{L}},  \qquad \bar{z} \rightarrow \bar{\omega}=e^{\frac{i2\pi \bar{z}}{L}}.
	\end{align}
	Under a general conformal transformation, the four-point correlator transforms as
\begin{align}\label{eq_4pf_trans}
\left<  \sigma_{n_o} (\omega_1,\bar{\omega}_1) \bar{\sigma}_{n_o} (\omega_2,\bar{\omega}_2) \bar{\sigma}_{n_o} (\omega_3,\bar{\omega}_3) \sigma_{n_o} (\omega_4,\bar{\omega}_4)  \right>=&\prod_{i=1}^{4}\left(\frac{d\omega}{d z}\right)^{-h_{n_o}}_{\omega=\omega_i}\left(\frac{d\bar{\omega}}{d \bar{z}}\right)^{-\bar{h}_{n_o}}_{\bar{\omega}=\bar{\omega}_i}\notag \\& \hspace{-2cm}\times\left<  \sigma_{n_o} (z_1,\bar{z}_1) \bar{\sigma}_{n_o} (z_2,\bar{z}_2) \bar{\sigma}_{n_o} (z_3,\bar{z}_3) \sigma_{n_o} (z_4,\bar{z}_4)  \right>.
\end{align}
Following a similar prescription as in \cref{sssec_nr_cft_dis_zt} along with \cref{eq_nr_fstrans,eq_4pf_trans}, the time dependent OEE in the large $c$ limit may be obtained as
	\begin{align}\label{eq_nr_cft_dis_fs_oee}
		S_o(A_1:A_2) & =  \frac{c}{12} \log \left[ \frac{1+ \sqrt{\zeta}}{1- \sqrt{\zeta}} \right] +\frac{c}{12}\log \left[ \frac{1+ \sqrt{\bar{\zeta}}}{1- \sqrt{\bar{\zeta}}} \right] +\frac{c}{3} \log \left[ \frac{L}{2\pi a} \sqrt{2\cos \frac{2\pi x_{41}}{L} -2\cos \frac{2\pi t_{41}}{L}}  \right] \notag \\ & + \frac{c}{3} \log \left[ \frac{L}{2\pi a} \sqrt{2\cos \frac{2\pi x_{32}}{L} -2\cos \frac{2\pi t_{32}}{L}}  \right]-\frac{c}{6} \log 2 , 
	\end{align}
	where the cross ratios are $\zeta=\frac{(\omega_2-\omega_1)(\omega_4-\omega_3)}{(\omega_3-\omega_1)(\omega_4-\omega_2)}$ and $\bar{\zeta}=\frac{(\bar{\omega}_2-\bar{\omega}_1)(\bar{\omega}_4-\bar{\omega}_3)}{(\bar{\omega}_3-\bar{\omega}_1)(\bar{\omega}_4-\bar{\omega}_2)}$ and $a$ is the UV cut-off as earlier. Note that the constant term arises from the conformal block expansion and there may be other undetermined OPE constants as earlier.

	\subsubsection{Two disjoint intervals at a finite temperature}\label{sssec_nr_cft_dis_ft}

A CFT$_2$ at a finite temperature $T=1/\beta$ is described on a cylinder of circumference $\beta$ along the Euclidean time direction. This may be obtained through the conformal map \cite{Calabrese:2009qy, Calabrese:2012nk}
	\begin{align}\label{eq_nr_fttrans}
		z \rightarrow \omega=e^{\frac{2\pi z}{\beta}}, \qquad \bar{z} \rightarrow \bar{\omega}=e^{\frac{2\pi \bar{z}}{\beta}}.
	\end{align} 
Following the prescription from \cref{sssec_nr_cft_dis_fs} and \cref{eq_nr_fttrans}, it is possible to compute the OEE for the mixed state configuration of two boosted disjoint intervals at a finite temperature for large central charge as
\begin{align}\label{eq_nr_cft_dis_ft_oee}
		S_o(A_1:A_2) & =  \frac{c}{12} \log \left[ \frac{1+ \sqrt{\xi}}{1- \sqrt{\xi}}\; \right] +\frac{c}{12}\log \left[ \frac{1+ \sqrt{\bar{\xi}}}{1- \sqrt{\bar{\xi}}}\; \right]  +\frac{c}{3} \log \left[ \frac{\beta}{2\pi a} \sqrt{2\cosh \frac{2\pi x_{41}}{\beta} -2\cosh \frac{2\pi t_{41}}{\beta}} \; \right] \notag
		\\ &   + \frac{c}{3} \log \left[ \frac{\beta}{2\pi a} \sqrt{2\cosh \frac{2\pi x_{32}}{\beta} -2\cosh \frac{2\pi t_{32}}{\beta}} \; \right] -\frac{c}{6} \log 2, 
	\end{align}
	where $\xi=\frac{(\omega_2-\omega_1)(\omega_4-\omega_3)}{(\omega_3-\omega_1)(\omega_4-\omega_2)}$ and $\bar{\xi}=\frac{(\bar{\omega}_2-\bar{\omega}_1)(\bar{\omega}_4-\bar{\omega}_3)}{(\bar{\omega}_3-\bar{\omega}_1)(\bar{\omega}_4-\bar{\omega}_2)}$ are the cross ratios, and $a$ is the UV cut-off. As earlier the constant term shown arises from the conformal block expansion while there may be other undetermined OPE constants.

	\subsection{OEE for two adjacent intervals}\label{ssec_nr_cft_adj}
	
	We now turn our attention to the computation of the time dependent OEE for bipartite mixed states described by two boosted adjacent intervals in CFT$_2$s.

	\subsubsection{Two adjacent intervals at zero temperature}\label{sssec_nr_cft_adj_zt}
	
In this case we consider two boosted adjacent intervals $A_1=[z_1,z_2]$ and $A_2=[z_2,z_3]$ in a CFT$_2$ at zero temperature. This configuration may be obtained from \cref{eq_4pf} by simply considering the limit $z_3 \to z_2$, and relabeling $z_4$ as $z_3$. The relevant twist correlator in this case is the following three-point function
	\begin{align} \label{eq_nr_3pf}
		\text{Tr} \left( \rho_{A_1A_2}^{T_{A_2}} \right)^{n_o}&= \left<  \sigma_{n_o} (z_1,\bar{z}_1) \bar{\sigma}_{n_o}^2 (z_2,\bar{z}_2) \sigma_{n_o} (z_3,\bar{z}_3)  \right>  \notag
		\\&= \frac{C}{((z_2-z_1)(z_3-z_1)(z_3-z_2))^{h_{n_o}} \ \ ((\bar{z}_2-\bar{z}_1)(\bar{z}_3-\bar{z}_1)(\bar{z}_3-\bar{z}_2))^{\bar{h}_{n_o}}}\ \ ,
	\end{align}
	where $C$ is the appropriate OPE coefficient.
	Using \cref{eq_vn_oee,eq_nr_3pf}, the time dependent OEE for two adjacent intervals at zero temperature in the large central charge limit may be calculated as
	\begin{align}\label{eq_nr_cft_adj_zt_oee}
		S_o(A_1:A_2)  = \frac{c}{6} \log \left[ \frac {\sqrt{x_{21}^2-t_{21}^2} }{a}\; \right]  +\frac{c}{6} \log \left[ \frac{ \sqrt{x_{32}^2-t_{32}^2} }{a}\; \right]+\frac{c}{6} \log \left[ \frac{ \sqrt{x_{31}^2-t_{31}^2} }{a}\; \right] + \text {constant} ,
	\end{align}
	where $a$ is again the UV cut-off and the constant arises from the OPE coefficient in \cref {eq_nr_3pf}.

	\subsubsection{Two adjacent intervals in a  finite size system}\label{sssec_nr_cft_adj_fs}
	
	Following the prescription in \cref{sssec_nr_cft_dis_fs}, we consider two boosted adjacent intervals on a cylinder of circumference $L$ [see the conformal map in \cref{eq_nr_fstrans}]. Constructing the relevant three-point twist correlator through \cref{eq_nr_fstrans,eq_nr_3pf}, the OEE in the large $c$ limit may now be obtained as
	\begin{align}\label{eq_nr_cft_adj_fs_oee}
		S_o(A_1:A_2) & = \frac{c}{6}  \log \left[ \frac{L}{2\pi a} \sqrt{2\cos \frac{2\pi x_{21}}{L} -2\cos \frac{2\pi t_{21}}{L}} \;\right] + \frac{c}{6}\log \left[ \frac{L}{2\pi a} \sqrt{2\cos \frac{2\pi x_{32}}{L} -2\cos \frac{2\pi t_{32}}{L}}\; \right] \notag \\ &  + \frac{c}{6} \log \left[ \frac{L}{2\pi a} \sqrt{2\cos \frac{2\pi x_{31}}{L} -2\cos \frac{2\pi t_{31}}{L}}\; \right] + \text {constant} .
	\end{align}

	\subsubsection{Two adjacent intervals at a finite temperature}\label{sssec_nr_cft_adj_ft}

	The relevant three-point twist correlator for the configuration of two boosted adjacent intervals on a cylinder of circumference $\beta$ may be constructed using \cref{eq_nr_fttrans,eq_nr_3pf}. The OEE for such a configuration in the large central charge limit may be computed as
	\begin{align}\label{eq_nr_cft_adj_ft_oee}
		S_o(A_1:A_2) & =  \frac{c}{6}  \log \left[ \frac{\beta}{2\pi a} \sqrt{2\cosh \frac{2\pi x_{21}}{\beta} -2\cosh \frac{2\pi t_{21}}{\beta}} \;\right] \notag \\
		& + \frac{c}{6} \log \left[ \frac{\beta}{2\pi a} \sqrt{2\cosh \frac{2\pi x_{32}}{\beta} -2\cosh \frac{2\pi t_{32}}{\beta}} \;\right]\notag
		\\&     + \frac{c}{6} \log \left[ \frac{\beta}{2\pi a} \sqrt{2\cosh \frac{2\pi x_{31}}{\beta} -2\cosh \frac{2\pi t_{31}}{\beta}}\; \right]+ \text {constant}. 
	\end{align}

	\subsection{OEE for a single interval}\label{ssec_nr_cft_sin}
	
	In this subsection we compute the OEE for bipartite states described by a boosted single interval in CFT$_2$s.
	
	\subsubsection{A single interval at zero temperature}\label{sssec_nr_cft_sin_zt}
	
	We begin with a boosted single interval $A_1=[z_1,z_2]$ in a CFT$_2$ at zero temperature. Such a configuration may be obtained from \cref{eq_4pf} through the limit $z_3 \to z_2$ and $z_4 \to z_1$. In this limit, the full infinite system is entirely described by $A_1 \cup A_2$ such that $A_2=A_1^c$. For this setup the four-point twist correlator in \cref{eq_4pf} reduces to the following two-point function
	\begin{align}\label{eq_nr_2pf}
		\text{Tr} \left( \rho_{A }^{T_2}\right)^{n_o}=\left< \sigma _{n_o}^2 (z_1,\bar{z_1})\bar{\sigma}_{n_o}^2 (z_2,\bar{z_2}) \right> .
	\end{align}

	Here the twist operator $ \sigma _{n_o}^2 $ connects the  $i$-th sheet with the $ (i+2) $-th sheet and have dimensions $ h_{n_o}^{(2)}=h_{n_o} $. As $n_o$ is odd, the twist correlator simply results in the reorganization of the replica sheets but does not change the structure of the $n_o$-sheeted Riemann surface.\footnote {Note that this is in contrast with even $n$ where the original Riemann surface decouples into two independent 
	Riemann surfaces as described in \cite{Calabrese:2012nk}.} As a result, we have
	\begin{align}
		\text{Tr}\left( \rho_{A}^{T_2} \right)^{n_o}=\left< \sigma _{n_o} (z_1,\bar{z_1})\bar{\sigma}_{n_o} (z_2,\bar{z_2}) \right>=\text{Tr} (\rho_{A})^{n_o}.
	\end{align}
	Using the two-point function
	\begin{align}\label{eq_2pf}
		\left< \sigma _{n_o} (z_1,\bar{z_1})\bar{\sigma}_{n_o} (z_2,\bar{z_2}) \right>=\frac{1}{(z_2-z_1)^{2h_{n_o}}  (\bar{z}_2-\bar{z}_1)^{2\bar{h}_{n_o}}},
	\end{align}
	it is possible to compute the time dependent OEE for a single interval at zero temperature at large $c$ as
	\begin{align}\label{eq_nr_cft_sin_zt_oee}
		S_o(A_1:A_2)  = \frac{c}{3} \log \left[ \frac {\sqrt{x_{21}^2-t_{21}^2} }{a}\; \right] ,
	\end{align}
	where $a$ is the UV cut-off.

	\subsubsection{A single interval in a finite size system}\label{sssec_nr_cft_sin_fs}

	 For the configuration of a single interval on a cylinder with circumference $L$, the relevant twist correlator may be constructed using the conformal map in \cref{eq_nr_fstrans} and the two-point function in \cref{eq_nr_2pf}. The time dependent OEE in the large $c$ approximation may thus be computed as 
	\begin{align}\label{eq_nr_cft_sin_fs_oee}
		S_o(A_1:A_2)= \frac{c}{3} \log \left[ \frac{L}{2\pi a} \sqrt{2\cos \frac{2\pi x_{21}}{L} -2\cos \frac{2\pi t_{21}}{L}} \right].
	\end{align}

	\subsubsection{A single interval at a finite temperature}\label{sssec_nr_cft_sin_ft}

	For the case of a single interval at a finite temperature on a thermal cylinder with circumference $\beta$ it is required to follow the construction described in \cite{Calabrese:2014yza}, for the partially transposed reduced density matrix in the context of entanglement negativity. To this end, for a boosted single interval $A=[z_2,z_3]$ it is necessary to consider two large but finite boosted auxiliary intervals $B_1=[z_1,z_2]$ and $B_2=[z_3,z_4]$, where $ z_1=-L+i \tau _1, z_2=-\ell +i \tau _2, z_3=i\tau _3, z_4=L+i\tau _4$ on either side of the interval $A$ describing a tripartite configuration. For this configuration the time dependent OEE in a bipartite limit $B \equiv B_1 \cup B_2 \to A^c$  $(L \to \infty)$ may be obtained as
	\begin{align}
		S_o(A:B)= \lim_{L \to \infty}  \lim_{n_o \to 1}  \frac{1}{1-n_o} \log  \left<\sigma_{n_o}(z_1,\bar{z}_1)\bar{\sigma}_{n_o}^2(z_2,\bar{z}_2)\sigma_{n_o}^2(z_3,\bar{z}_3)\bar{\sigma}_{n_o}(z_4,\bar{z}_4)\right> .
	\end{align}	
	Note that the bipartite limit  must be implemented after the replica limit $n_o \to 1$ in the above equation.\footnote{See \cite{Calabrese:2014yza} for a detailed explanation.} On the complex plane, the relevant four-point twist correlator may be expressed as
	\begin{align}\label{eq_sin_ft_4pf}
		& \left<\sigma_{n_o}(z_1,\bar{z}_1)\bar{\sigma}_{n_o}^2(z_2,\bar{z}_2)\sigma_{n_o}^2(z_3,\bar{z}_3)\bar{\sigma}_{n_o}(z_4,\bar{z}_4)\right> \notag \\
		& \qquad \qquad \qquad=  \frac{\mathcal{F}_{n_o}(\eta)}{\left((z_4-z_1)(z_3-z_2)\right)^{2h_{n_o}} \eta^{h_{n_o}}} \frac{\mathcal{\bar{F}}_{n_o}(\bar{\eta})}{\left((\bar{z}_4-\bar{z}_1)(\bar{z}_3-\bar{z}_2)\right)^{2\bar{h}_{n_o}} \bar{\eta}^{\bar{h}_{n_o}}} ,
	\end{align}	
	where $\eta$ and $\bar{\eta}$ are the cross ratios as defined in \cref{sssec_nr_cft_dis_zt}, and $\mathcal{F}_{n_o}(\eta)$ and $\bar{\mathcal{F}}_{n_o}(\bar{\eta})$ are arbitrary non-universal functions of the cross ratios. These non-universal functions are given in the limits $\eta \to 1$ and $\eta \to 0$ as \cite{Basak:2022gcv}
	\begin{align}
		\mathcal{F}_{n_o} (1)=1 \ \ , \qquad \qquad \mathcal{F}_{n_o} (0)=C_{n_o}.
	\end{align}
	Here $C_{n_o}$ is a non-universal constant which depends on the full operator content of the field theory. The relevant four-point correlator for the configuration described above may be obtained from the map given in \cref{eq_nr_fttrans}. Defining the non-universal functions $f(\xi)$ and $\bar{f}(\bar{\xi})$ such that
	\begin{align}
	f(\xi)=\lim_{n_o \to 1} \log [\mathcal{F}_{n_o}(\xi)], \qquad \bar{f}(\bar{\xi})=\lim_{n_o \to 1} \log [\mathcal{\bar{F}}_{n_o}(\bar{\xi})],
	\end{align}		
	 the OEE for a single interval in a thermal CFT$_2$ at large $c$ may be computed as	
	\begin{align}\label{eq_nr_cft_sin_ft_oee}
		S_o(A:A^c) & = \lim_{L \to \infty} \; \frac{c}{3} \log \left[ \frac{\beta}{2\pi a} \sqrt{2\cosh \frac{4\pi L}{\beta} -2\cosh \frac{2\pi t_{41}}{\beta}} \; \right] \notag \\ &  +\frac{c}{3} \log \left[ \frac{\beta}{2\pi a} \sqrt{2\cosh \frac{2\pi \ell}{\beta} -2\cosh \frac{2\pi t_{32}}{\beta}} \; \right]
		-\frac{c \pi \ell}{3 \beta} +f \left( e^{\frac{2 \pi }{\beta}(-\ell+t_{32})} \right)+\bar{f} \left( e^{\frac{2 \pi }{\beta}(-\ell-t_{32})} \right),
	\end{align}	
	where $\xi$ and $\bar{\xi}$ are cross ratios on the thermal cylinder as defined in \cref{sssec_nr_cft_dis_ft}. Note that in the large $ c $ limit the non-universal contributions described by the functions $ f $ and $ \bar{f} $ are suppressed.

	\section{Covariant Holographic OEE in CFT$_2$s }\label{sec_nr_holo}
	
	We now turn our attention to a brief review  of the holographic description for the odd entanglement entropy of bipartite mixed states considered earlier in CFT$_2$s. The bulk dual AdS$_3$ spacetime may be embedded in flat $\mathbb{R}^{(2,2)}$ as follows \cite{Kusuki:2019evw}
	\begin{align}\label{eq_minkmetric}
		\text{d}s^2= \eta_{AB} \text{d}X^A \text{d}X^B = & -\text{d}T_1^2-\text{d}T_2^2+\text{d}X_1^2+\text{d}X_2^2, \\
		-T_1^2-T_2^2+&X_1^2+X_2^2=-1.
	\end{align}	
	The dual bulk geometry for the zero temperature state in a CFT$_2$ is described by a pure AdS$ _3 $ spacetime in Poincar\'e coordinates as follows
	\begin{align}\label{eq_pmetric}
		\text{d}s^2 = \frac{1}{z^2} (-\text{d}t^2+\text{d}x^2+\text{d}z^2),
	\end{align}
	where the AdS radius has been set to unity. The metric in \cref{eq_pmetric} may be mapped to that in \cref{eq_minkmetric} through  the following embedding relations
	\begin{align}\label{eq_pemb}
		T_1=\frac{t}{z}, \qquad & \qquad T_2= \frac{1-t^2+x^2+z^2}{2z}, \notag \\ X_1=\frac{x}{z}, \qquad & \qquad X_2=\frac{1+t^2-x^2-z^2}{2z}.
	\end{align}
Note that the conformal boundary is located at $z=a$, where $a$ is the UV cut-off of the dual CFT$ _2 $. 
	
	For a zero temperature CFT$_2$ of finite size $L$, the bulk dual is a pure $\text{AdS}_3$ spacetime in global coordinates given as \cite{Ryu:2006ef}
	\begin{align}\label{eq_gmetric}
		\text{d}s^2 =   -\cosh ^2 \rho \text{d}\tau^2 + \text{d}\rho^2 + \sinh ^2 \rho \text{d}\phi^2 ,
	\end{align}
	where $\tau =2 \pi t/L  $ and $\phi =2 \pi x/L$. 
	The corresponding embedding relations in this case are described as follows
	\begin{align}\label{eq_gemb}
		T_1= \cosh \rho \sin \frac{2 \pi t}{L}, \qquad & \qquad T_2= \cosh \rho \cos \frac{2 \pi t}{L}, \notag \\ X_1= \sinh \rho \cos \frac{2 \pi x}{L}, \qquad & \qquad X_2= \sinh \rho \sin \frac{2 \pi x}{L}.
	\end{align}
The conformal boundary for the metric described in \cref{eq_gmetric} lies at $\rho=\rho_c$. 
Observe that the bulk IR cut-off $\rho_c$ is related to the UV cut-off of the dual CFT as $e^{\rho_c} \sim 1/a$.
	
	Finally the bulk dual geometry for a finite temperature state in a CFT$_2$ is characterized by a BTZ black hole geometry represented by the following metric \cite{Carlip:1995qv}
	\begin{align}\label{eq_btzmetric}
		\text{d}s^2 = -(r^2-r_h^2)\text{d}t^2 + \frac{1}{(r^2-r_h^2)}\text{d}r^2 + r^2\text{d}x^2,
	\end{align}
	where $ 1/\beta $ is the temperature and the horizon of the black hole is located at 
	$r=r_h\equiv 2 \pi /\beta$. The metric in the BTZ coordinates in \cref{eq_btzmetric} may be mapped to that in the embedding coordinates through the following relations \cite{Carlip:1995qv,Shenker:2013pqa}
	\begin{align}\label{eq_btzemb}
		T_1=\sqrt{\left(\frac{r^2}{r_h^2} -1\right)} \sinh \left( \frac{2\pi t}{\beta} \right), \qquad & \qquad T_2=\frac{r}{r_h} \cosh \left( \frac{2\pi x}{\beta} \right), \notag \\ X_1=\sqrt{\left(\frac{r^2}{r_h^2} -1\right)} \cosh \left( \frac{2\pi t}{\beta} \right), \qquad & \qquad X_2=\frac{r}{r_h} \sinh \left( \frac{2\pi x}{\beta} \right).
	\end{align}
	The conformal boundary for the BTZ spacetime in \cref{eq_btzmetric} is  situated at $r=r_c$, where the bulk IR cut-off $r_c$ is related to the corresponding UV cut-off of the dual CFT as $r_c \sim 1/a$. 
	
	It is well known that the entanglement entropy for a single interval in the dual CFT$_2$ with end points $X(s_i)$ and $X(s_j)$ (in the embedding coordinates), is described by \cite{Ryu:2006bv,Hubeny:2007xt}
	\begin{align}\label{eq_holo_ee}
		S=\frac{\Delta s}{4 G_N},
	\end{align}
	where the bulk geodesic length $\Delta s$ between the points $X(s_i)$ and $X(s_j)$ is given as \cite {Kusuki:2019evw}
	\begin{align}\label{eq_geoleng}
		\Delta s = \cosh ^{-1} [- X(s_i) \cdot X(s_j) ]\ \ .
	\end{align}
	Note that $s_j$ represents the standard coordinates in the dual bulk AdS$_3$ geometry.

	\subsection{Holographic OEE for two disjoint intervals}\label{ssec_nr_holo_dis}

	We are now in a position to describe the computation for the EWCS of two boosted disjoint intervals $A_1=[(x_1,t_1),(x_2,t_2)]$ and $A_2=[(x_3,t_3),(x_4,t_4)]$ as discussed earlier in \cref{sec_nr_cft}. The corresponding end points in the bulk embedding coordinates are given as $X(s_j)$ for $j=1,2,3,4$. The corresponding EWCS involving four arbitrary bulk points may be described as \cite{Kusuki:2019evw,Boruch:2020wbe}
	\begin{align}\label{eq_nr_dis_ew}
		E_W(A_1:A_2)=\frac{1}{4G_N} \cosh ^{-1} \left[ \frac{1+\sqrt{u}}{\sqrt{v}}\; \right],
	\end{align}
	where $u$ and $v$ are given as
	\begin{align}
		u=\frac{\xi_{12}^{-1} \xi_{34}^{-1}}{\xi_{13}^{-1} \xi_{24}^{-1}},\  \ v=\frac{\xi_{14}^{-1} \xi_{23}^{-1}}{\xi_{13}^{-1} \xi_{24}^{-1}},  \qquad\qquad \xi_{ij}^{-1}=-X(s_i) \cdot X(s_j) .
	\end{align}

	\subsubsection{Two disjoint intervals at zero temperature}\label{sssec_nr_holo_dis_zt}
	
The EWCS for a zero temperature state described by the two boosted disjoint intervals involving the bulk points $X(s_1), X(s_2), X(s_3)$ and $X(s_4)$ may be computed from \cref{eq_nr_dis_ew} as follows
\begin{align}\label{eq_nr_dis_ew_zt}
	E_W(A_1:A_2)=\frac{c}{12}\log \left[ \frac{1+\sqrt{\eta}}{1-\sqrt{\eta}} \right] + \frac{c}{12}\log \left[ \frac{1+\sqrt{\bar{\eta}}}{1-\sqrt{\bar{\eta}}}\right],
\end{align}
	where
	\begin{align}
		\eta=\frac{(x_{21}+t_{21}) (x_{43}+t_{43})}{(x_{31}+t_{31} (x_{42}+t_{42}))}, \qquad \bar{\eta}=\frac{(x_{21}-t_{21}) (x_{43}-t_{43})}{(x_{31}-t_{31} (x_{42}-t_{42}))},
	\end{align}
	are the cross ratios previously defined in \cref{sssec_nr_cft_dis_zt}, and the Brown Henneaux formula \cite{Brown:1986nw} has been utilized. The relevant holographic entropy may be obtained from \cref{eq_holo_ee} as
	\begin{align}\label{eq_nr_cft_dis_zt_ee}
		S(A_1 \cup A_2)=\frac{c}{3} \log \left[ \frac {\sqrt{x_{41}^2-t_{41}^2} }{a} \right]  +\frac{c}{3} \log \left[ \frac{ \sqrt{x_{32}^2-t_{32}^2} }{a} \right].
	\end{align}
	The holographic OEE may now be obtained by substituting \cref{eq_nr_dis_ew_zt,eq_nr_cft_dis_zt_ee} in the relation described in \cref{eq_holodual}. Interestingly our holographic result matches with the field theory replica technique computation described in \cref{eq_nr_cft_dis_zt_oee} in the large $c$ limit modulo certain constants. This serves as a strong consistency check for our holographic construction.

	\subsubsection{Two disjoint intervals in a finite size system}\label{sssec_nr_holo_dis_fs}

In this case for a given set of embedding coordinates $X(s_j)$ (where $j=1,2,3,4$) representing end points in global AdS spacetime, the EWCS for two boosted disjoint intervals in a zero temperature finite size system may be computed as
	\begin{align}\label{eq_nr_dis_ew_fs}
		E_W(A_1:A_2)=\frac{c}{12}\log \left[ \frac{1+\sqrt{\zeta}}{1-\sqrt{\zeta}} \right] + \frac{c}{12}\log \left[ \frac{1+\sqrt{\bar{\zeta}}}{1-\sqrt{\bar{\zeta}}} \right],
	\end{align}
	where
	\begin{align}
		\zeta=\frac{\sinh (\frac{\pi (x_{21}+t_{21})}{L}) \sinh (\frac{\pi (x_{43}+t_{43})}{L})}{\sinh (\frac{\pi (x_{31}+t_{31})}{L}) \sinh (\frac{\pi (x_{42}+t_{42})}{L})}, \qquad \bar{\zeta}=\frac{\sinh (\frac{\pi (x_{21}-t_{21})}{L}) \sinh (\frac{\pi (x_{43}-t_{43})}{L})}{\sinh (\frac{\pi (x_{31}-t_{31})}{L}) \sinh (\frac{\pi (x_{42}-t_{42})}{L})},
	\end{align}
	are the cross ratios as defined in \cref{sssec_nr_cft_dis_fs}. The holographic entanglement entropy may then be obtained using \cref{eq_holo_ee} as
	\begin{align}\label{eq_nr_cft_dis_fs_ee}
		S(A_1 \cup A_2) & = \frac{c}{3} \log \left[ \frac{L}{2\pi a} \sqrt{2\cos \frac{2\pi x_{41}}{L} -2\cos \frac{2\pi t_{41}}{L}} \; \right] \notag \\ &  + \frac{c}{3} \log \left[ \frac{L}{2\pi a} \sqrt{2\cos \frac{2\pi x_{32}}{L} -2\cos \frac{2\pi t_{32}}{L}} \; \right].
	\end{align}
	It is now straightforward to obtain the holographic OEE using \cref{eq_holodual,eq_nr_dis_ew_fs,eq_nr_cft_dis_fs_ee}, which matches with that obtained from the replica technique in \cref{eq_nr_cft_dis_fs_oee} at large $c$ modulo constants. This again strongly supports our holographic construction.

	\subsubsection{Two disjoint intervals at a finite temperature}\label{sssec_nr_holo_dis_ft}
	
	For two boosted disjoint intervals at a finite temperature, the EWCS  may be computed for the embedding coordinates $X(s_1), X(s_2), X(s_3)$ and $X(s_4)$ by employing \cref{eq_nr_dis_ew} as follows
	\begin{align}\label{eq_nr_dis_ew_ft}
		E_W(A_1:A_2)=\frac{c}{12}\log \left[ \frac{1+\sqrt{\xi}}{1-\sqrt{\xi}} \right] + \frac{c}{12}\log \left[ \frac{1+\sqrt{\bar{\xi}}}{1-\sqrt{\bar{\xi}}} \right],
	\end{align}
	where 
	\begin{align}
		\xi=\frac{\sin (\frac{\pi (x_{21}+t_{21})}{\beta}) \sin (\frac{\pi (x_{43}+t_{43})}{\beta})}{\sin (\frac{\pi (x_{31}+t_{31})}{\beta}) \sin (\frac{\pi (x_{42}+t_{42})}{\beta})}, \qquad \bar{\xi}=\frac{\sin (\frac{\pi (x_{21}-t_{21})}{\beta}) \sin (\frac{\pi (x_{43}-t_{43})}{\beta})}{\sin (\frac{\pi (x_{31}-t_{31})}{\beta}) \sin (\frac{\pi (x_{42}-t_{42})}{\beta})},
	\end{align}	
	are the cross ratios as defined in \cref{sssec_nr_cft_dis_ft}. The holographic entanglement entropy may be obtained using \cref{eq_holo_ee} as 
	\begin{align}\label{eq_nr_cft_dis_ft_ee}
		S(A_1 \cup A_2) & = \frac{c}{3} \log \left[ \frac{\beta}{2\pi a} \sqrt{2\cosh \frac{2\pi x_{41}}{\beta} -2\cosh \frac{2\pi t_{41}}{\beta}} \; \right] \notag
		\\ &   + \frac{c}{3} \log \left[ \frac{\beta}{2\pi a} \sqrt{2\cosh \frac{2\pi x_{32}}{\beta} -2\cosh \frac{2\pi t_{32}}{\beta}} \; \right].
	\end{align}
Interestingly the holographic OEE computed through the relation in \cref{eq_holodual,eq_nr_dis_ew_ft,eq_nr_cft_dis_ft_ee} matches with the field theory calculations in \cref{eq_nr_cft_dis_ft_oee} in the large $c$ limit up to certain constants, which once again substantiates our construction.

	\subsection{Holographic OEE for two adjacent intervals}\label{ssec_nr_holo_adj}
	
	We now outline the procedure to compute the EWCS for two boosted adjacent intervals $A_1=[(x_1,t_1),(x_2,t_2)]$ and $A_2=[(x_2,t_2),(x_3,t_3)]$, where the corresponding end points in the bulk embedded coordinates may be represented as $X(s_j)$ with $j=1,2,3$. The EWCS for this configuration is given as \cite{Kusuki:2019evw,Basu:2023jtf}
	\begin{align}\label{eq_nr_adj_ew}
		E_W(A_1:A_2)=\frac{1}{4G_N} \cosh ^{-1} \left[ \frac{\sqrt{2}}{\sqrt{v}} \right],
	\end{align}
	where 
	\begin{align}
		v=\frac{\xi_{13}^{-1} }{\xi_{12}^{-1} \xi_{23}^{-1}} , \qquad \xi_{ij}^{-1}=-X(s_i) \cdot X(s_j) .
	\end{align}

	\subsubsection{Two adjacent intervals at zero temperature}\label{sssec_nr_holo_adj_zt}
	
	Using a prescription similar to that described in \cref{ssec_nr_holo_dis}, the EWCS for two boosted adjacent intervals at zero temperature may be computed from \cref{eq_nr_adj_ew} as follows
	\begin{align}\label{eq_nr_adj_ew_zt}
		E_W(A_1:A_2) & =  \frac{1}{4G_N} \cosh^{-1} \left[ \frac{\sqrt{x_{21}^2-t_{21}^2}\sqrt{x_{32}^2-t_{32}^2}}{a\sqrt{x_{31}^2-t_{31}^2}} \right] \notag \\
		&= \frac{c}{6} \log \left[ \frac{\sqrt{x_{32}^2-t_{32}^2}}{a} \right] +\frac{c}{6} \log \left[ \frac{\sqrt{x_{21}^2-t_{21}^2}}{a} \right]  - \frac{c}{6} \log \left[ \frac{\sqrt{x_{31}^2-t_{31}^2}}{a} \right] +\frac{c}{6} \log 2.
	\end{align} 
	The relevant HEE may be obtained as earlier from \cref{eq_holo_ee} as follows
	\begin{align}\label{eq_nr_cft_adj_zt_ee}
		S(A_1 \cup A_2)=\frac{c}{3} \log \left[ \frac{ \sqrt{x_{31}^2-t_{31}^2} }{a} \right] .
	\end{align}
	We may now obtain the holographic OEE for the two subsystems $ A_1 $ and $ A_2 $ from \cref{eq_holodual,eq_nr_adj_ew_zt,eq_nr_cft_adj_zt_ee}. Our holographic result reproduces the corresponding replica technique calculation at large $ c $ as given by \cref{eq_nr_cft_adj_zt_oee}, modulo certain constant terms. This once again serves to verify the holographic conjecture.

	\subsubsection{Two adjacent intervals in a finite size system}\label{sssec_nr_holo_adj_fs}
	
	Using \cref{eq_nr_adj_ew}, the EWCS for two boosted adjacent intervals for a finite size system at zero temperature may be calculated as
	\begin{align}\label{eq_nr_adj_ew_fs}
		E_W(A_1:A_2) & = \frac{1}{4G_N} \cosh ^{-1} \left[ \sqrt{2} \frac{L}{2 \pi a}\frac{ \sqrt{ \cos \left( \frac{2 \pi x_{21}}{L} \right) -  \cos \left( \frac{2 \pi t_{21}}{L} \right)} \sqrt{ \cos \left( \frac{2 \pi x_{32}}{L} \right) -  \cos \left( \frac{2 \pi t_{32}}{L} \right)}}{\sqrt{ \cos \left( \frac{2 \pi x_{31}}{L} \right) -  \cos \left( \frac{2 \pi t_{31}}{L} \right)}}   \;\right] \notag \\
		& =  \frac{c}{6} \log \left[ \frac{L}{2 \pi a} \sqrt{2 \cos \left( \frac{2 \pi x_{21}}{L} \right) - 2 \cos \left( \frac{2 \pi t_{21}}{L} \right)}\; \right] \notag \\
		& +\frac{c}{6} \log \left[ \frac{L}{2 \pi a} \sqrt{2 \cos \left( \frac{2 \pi x_{32}}{L} \right) - 2 \cos \left( \frac{2 \pi t_{32}}{L} \right)}\; \right] \notag \\ 
		&  - \frac{c}{6} \log \left[ \frac{L}{2 \pi a} \sqrt{2 \cos \left( \frac{2 \pi x_{31}}{L} \right) - 2 \cos \left( \frac{2 \pi t_{31}}{L} \right)}\; \right] +\frac{c}{6} \log 2.
	\end{align}
	The quantity $S(A_1 \cup A_2)$  may be obtained using \cref{eq_holo_ee} as follows
	\begin{align}\label{eq_nr_cft_adj_fs_ee}
		S(A_1 \cup A_2)=\frac{c}{3} \log \left[ \frac{L}{2\pi a} \sqrt{2\cos \frac{2\pi x_{31}}{L} -2\cos \frac{2\pi t_{31}}{L}}\; \right].
	\end{align}
	It is now possible to obtain the holographic OEE using \cref{eq_holodual,eq_nr_adj_ew_fs,eq_nr_cft_adj_fs_ee}. Once again it is observed that this holographic result matches with the large $c$ field theory computation in \cref{eq_nr_cft_adj_fs_oee} modulo constant terms. This reinforces the consistency of our holographic proposal.

	\subsubsection{Two adjacent intervals at a finite temperature}\label{sssec_nr_holo_adj_ft}
	
	The EWCS for two boosted adjacent intervals for a system at a finite temperature may be computed using \cref{eq_nr_adj_ew} as follows
	\begin{align}\label{eq_nr_adj_ew_ft}
		E_W & =\frac{1}{4G_N} \cosh ^{-1} \left[ \sqrt{2} \frac{\beta}{2 \pi a}\frac{ \sqrt{ \cosh \left( \frac{2 \pi x_{21}}{\beta} \right) -  \cosh \left( \frac{2 \pi t_{21}}{\beta} \right)} \sqrt{ \cosh \left( \frac{2 \pi x_{32}}{\beta} \right) -  \cosh \left( \frac{2 \pi t_{32}}{\beta} \right)}}{\sqrt{ \cosh \left( \frac{2 \pi x_{31}}{\beta} \right) -  \cosh \left( \frac{2 \pi t_{31}}{\beta} \right)}}   \;\right] \notag \\
		& =  \frac{c}{6} \log \left[ \frac{\beta}{2 \pi a} \sqrt{2 \cosh \left( \frac{2 \pi x_{21}}{\beta} \right) - 2 \cosh \left( \frac{2 \pi t_{21}}{\beta} \right)}\; \right] \notag \\
		& +\frac{c}{6} \log \left[ \frac{\beta}{2 \pi a} \sqrt{2 \cosh \left( \frac{2 \pi x_{32}}{\beta} \right) - 2 \cosh \left( \frac{2 \pi t_{32}}{\beta} \right)} \;\right] \notag \\ 
		& - \frac{c}{6} \log \left[ \frac{\beta}{2 \pi a} \sqrt{2 \cosh \left( \frac{2 \pi x_{31}}{\beta} \right) - 2 \cosh \left( \frac{2 \pi t_{31}}{\beta} \right)}\; \right] +\frac{c}{6} \log 2.
	\end{align}
The holographic entanglement entropy  may be obtained using \cref{eq_holo_ee} as
	\begin{align}\label{eq_nr_cft_adj_ft_ee}
		S(A_1 \cup A_2)=\frac{c}{3} \log \left[ \frac{\beta}{2\pi a} \sqrt{2\cosh \frac{2\pi x_{31}}{\beta} -2\cosh \frac{2\pi t_{31}}{\beta}} \;\right].
	\end{align}
	As earlier the holographic OEE may be determined utilizing \cref{eq_holodual,eq_nr_adj_ew_ft,eq_nr_cft_adj_ft_ee}. Once again the holographic result matches with that obtained from the field theory computations in  \cref{eq_nr_cft_adj_ft_oee} in the large $c$ limit up to constant terms.

	\subsection{Holographic OEE for a single interval}\label{ssec_nr_holo_sin}

	In this subsection we compute the EWCS for a boosted single  interval $A=[(x_1,t_1),(x_2,t_2)]$, where $X(s_1)$ and $X(s_2)$ represent the end points in the embedding coordinates.

	\subsubsection{A single interval at zero temperature}\label{sssec_nr_holo_sin_zt}

The EWCS for a boosted single interval at zero temperature may be computed utilizing the following expression \cite{Kusuki:2019evw}
	\begin{align}\label{eq_nr_sin_ew}
		E_W=\frac{1}{4G_N} \cosh^{-1} \left[ -X(s_2) \cdot X(s_1) \right].
	\end{align}
	For the given configuration, the EWCS is given as
	\begin{align}\label{eq_nr_sin_ew_zt}
		E_W & =\frac{1}{4 G_N} \cosh ^{-1} \left[ \frac{\sqrt{x_{21}^2-t_{21}^2}}{2a} \right] \notag \\
		& =\frac{c}{3} \log \left[ \frac{\sqrt{x_{21}^2-x_{21}^2}}{a} \right].
	\end{align}
	
	Since the entire system $A \cup A^c$ is in a pure state, the holographic entanglement entropy $S(A \cup A^c)$ vanishes, and \cref{eq_holodual} suggests that the holographic OEE is identical to the EWCS. It is observed that as earlier the holographic result in \cref{eq_nr_sin_ew_zt} matches with the corresponding field theory computation in \cref{eq_nr_cft_sin_zt_oee}.

	\subsubsection{A single interval in a finite size system}\label{sssec_nr_holo_sin_fs}

	Using \cref{eq_nr_sin_ew}, the EWCS for a boosted single interval in a finite size system at zero temperature may be computed as
	\begin{align}\label{eq_nr_sin_ew_fs}
		E_W & =\frac{1}{4 G_N} \cosh ^{-1} \left[  \left(\frac{L}{2 \pi a}\right)^2 \left( \cos  \frac{2 \pi x_{21}}{L}  - \cos  \frac{2 \pi t_{21}}{L}  \right) \right] \notag \\
		& =\frac{c}{3} \log \left[ \frac{L}{2 \pi a} \sqrt{2 \cos \left( \frac{2 \pi x_{21}}{L} \right) - 2 \cos \left( \frac{2 \pi t_{21}}{L} \right)}\; \right].
	\end{align}
	
	As elaborated in \cref{sssec_nr_holo_sin_zt}, the holographic OEE  is identical to the EWCS  as the holographic entropy $S(A \cup A^c)$ vanishes. The holographic result is again shown to match with the CFT computation in \cref{eq_nr_cft_sin_fs_oee}.

	\subsubsection{A single interval at a finite temperature}\label{sssec_nr_holo_sin_ft}

Following the prescription described in \cref{sssec_nr_cft_sin_ft}, two finite but large boosted auxiliary intervals $B_1=[(x_1,t_1),(x_2,t_2)]$ and $B_2=[(x_3,t_3),(x_4,t_4)]$ are required to be considered on either side of the boosted interval in question, $A=[(x_2,t_2),(x_3,t_3)]$. In this case we may compute the quantity \cite{KumarBasak:2020eia,Basu:2021awn,Basu:2022nds}
	\begin{align}\label{eq_nr_sin_ew_n}
		\tilde{E}_W (A:B)=E_W(A:B_1)+E_W(A:B_2),
	\end{align} 
	where $\tilde{E}_W$ denotes an upper bound on the EWCS between the intervals $A$ and $B$, with $B\equiv B_1\cup B_2$. Note that each term on the right hand side of \cref{eq_nr_sin_ew_n} represents the EWCS of two adjacent intervals. Thus the results from \cref{sssec_nr_holo_adj_ft}  may be utilized to express these terms as
	\begin{align}\label{eq_nr_holo_sin_ew_ft_1}
		E_W(A:B_1) & =  \frac{c}{6} \log \left[ \frac{\beta}{2 \pi a} \sqrt{2 \cosh \left( \frac{2 \pi (L-\ell )}{\beta} \right) - 2 \cosh \left( \frac{2 \pi t_{21}}{\beta} \right)}\; \right] \notag \\ &  +\frac{c}{6} \log \left[ \frac{\beta}{2 \pi a} \sqrt{2 \cosh \left( \frac{2 \pi \ell}{\beta} \right) - 2 \cosh \left( \frac{2 \pi t_{32}}{\beta} \right)}\; \right] \notag \\ &  - \frac{c}{6} \log \left[ \frac{\beta}{2 \pi a} \sqrt{2 \cosh \left( \frac{2 \pi L}{\beta} \right) - 2 \cosh \left( \frac{2 \pi t_{31}}{\beta} \right)} \;\right] +\frac{c}{6} \log 2 ,
	\end{align}
	and
	\begin{align}\label{eq_nr_holo_sin_ew_ft_2}
		E_W(A:B_2) & = \frac{c}{6} \log \left[ \frac{\beta}{2 \pi a} \sqrt{2 \cosh \left( \frac{2 \pi \ell}{\beta} \right) - 2 \cosh \left( \frac{2 \pi t_{32}}{\beta} \right)}\; \right] \notag \\ &  +\frac{c}{6} \log \left[ \frac{\beta}{2 \pi a} \sqrt{2 \cosh \left( \frac{2 \pi L}{\beta} \right) - 2 \cosh \left( \frac{2 \pi t_{43}}{\beta} \right)}\; \right] \notag \\ & - \frac{c}{6} \log \left[ \frac{\beta}{2 \pi a} \sqrt{2 \cosh \left( \frac{2 \pi (L+\ell)}{\beta} \right) - 2 \cosh \left( \frac{2 \pi t_{42}}{\beta} \right)}\; \right] +\frac{c}{6} \log 2.
	\end{align}
	In accordance with \cref{eq_nr_sin_ew_n}, the upper bound for the EWCS between the subsystems $A$ and $B$ is given as
	\begin{align}
		\tilde{E}_W= & \frac{c}{6}  \log \left[\sqrt{\frac{\left(\cosh \left(\frac{2 \pi  L}{\beta }\right)-\cosh \left(\frac{2 \pi  t_{43}}{\beta }\right)\right) \left(\cosh \left(\frac{2 \pi  (L-l)}{\beta }\right)-\cosh \left(\frac{2 \pi  t_{21}}{\beta }\right)\right)}{\left(\cosh \left(\frac{2 \pi  L}{\beta }\right)-\cosh \left(\frac{2 \pi  t_{31}}{\beta }\right)\right) \left(\cosh \left(\frac{2 \pi  (l+L)}{\beta }\right)-\cosh \left(\frac{2 \pi  t_{42}}{\beta }\right)\right)}}\;\right] \notag \\ & \qquad +  \frac{c}{3}  \log \left[\frac{\beta }{2 \pi  a} \sqrt{2 \cosh \left(\frac{2 \pi  l}{\beta }\right)-2 \cosh \left(\frac{2 \pi  t_{32}}{\beta }\right)}\; \right] +\frac{c}{3} \log 2.
	\end{align}
	Implementing the bipartite limit $L \to \infty$ ($B\to A^c$) in the boundary, the above equation reduces to the EWCS for the single interval $A$ as follows
	\begin{align}\label{eq_nr_sin_ew_ft}
		\tilde{E}_W= & \frac{c}{3}  \log \left[\frac{\beta }{2 \pi  a} \sqrt{2 \cosh \left(\frac{2 \pi  l}{\beta }\right)-2 \cosh \left(\frac{2 \pi  t_{32}}{\beta }\right)} \;\right]-\frac{c \pi \ell}{3 \beta} +\frac{c}{3} \log 2.
	\end{align}
	The corresponding holographic entanglement entropy for the bipartite limit 
	 is given as \cite{Hubeny:2007xt} 
	\begin{align}\label{eq_nr_sin_hee_ft}
		S(A \cup A^c)=\lim_{L \to \infty} \left( \frac{c}{3} \log \left[ \frac{\beta}{2\pi a} \sqrt{2\cosh \frac{4\pi L}{\beta} -2\cosh \frac{2\pi t_{41}}{\beta}}\;  \right] \right).
	\end{align}
	Combining \cref{eq_nr_sin_ew_ft,eq_nr_sin_hee_ft} through the relation in \cref{eq_holodual} leads to the holographic OEE. We observe that the result matches (up to the constant term) with the OEE computed from field theory in \cref{eq_nr_cft_sin_ft_oee} at large $c$. Once again this reinforces the consistency of our holographic construction.

	\section{Time dependent OEE in CFT$_2$s with a conserved charge}\label{CFT_OEE_R_SYS}
	
In this section we describe the computation of the time dependent OEE in CFT$_2$s with a conserved charge, dual to non-extremal and extremal rotating BTZ black holes in the bulk. For a CFT$_2$ with a conserved charge at a finite temperature $1/\beta$, the Euclidean partition function may be described as\cite{Hubeny:2007xt,Caputa:2013lfa} 
	\begin{align}\label{ROT_PARTITION}
		Z(\beta) & =\text{Tr exp} [-\beta(H+i\Omega_E J)] \notag \\
		& =\text{Tr exp} [-\beta_+ L_0-\beta_- \bar{L}_0],
	\end{align}
	where we have $\beta_{\pm}=\beta\left(1\pm i\Omega_E\right)$. For the partition function described by \cref{ROT_PARTITION}, the angular momentum $J$ of the bulk rotating BTZ black hole leads to the conserved charge of the dual CFT$_2$. The Euclidean angular velocity of the black hole is given by $\Omega_E = - i \Omega$ and the Hamiltonian may be expressed as $H=L_0+\bar{L}_0$, where $L_0$ and $\bar{L}_0$ describe the standard Virasoro zero modes of the CFT$_2$. This thermal CFT$_2$ with a conserved charge may be described on a two dimensional ``twisted cylinder.''\footnote{The holomorphic (left moving) and anti-holomorphic (right moving) sectors are compactified with circumferences $\beta_+$ and $\beta_-$ respectively. These separate compactifications in turn give rise to two different effective inverse ``temperatures'' $\beta_+$ and $\beta_-$, also known as the left and right moving inverse temperatures respectively. See \cite{Caputa:2013eka} for details.} It is possible to map this twisted cylinder to the complex plane via the following conformal transformation
	\begin{align} \label{EQ_NE_TRN}
		z\rightarrow\omega=e^{\frac{2\pi}{\beta_+}z}, \qquad \bar{z}\rightarrow\bar{\omega}=e^{\frac{2\pi}{\beta_-}\bar{z}},
	\end{align}
	where $(z, \bar{z})$ and $(\omega, \bar{\omega})$ denote the coordinates on the complex plane and the twisted cylinder respectively. The bulk dual of this thermal CFT$_2$ is described by a non-extremal rotating BTZ black hole with temperature and angular velocity (potential) given as \cite{Caputa:2013lfa,Banados:1992wn,Banados:1992gq}
	\begin{align} \label{eq_r_cft_temp}
		T=\frac{1}{\beta}=\frac{r_+^2-r_-^2}{2 \pi r_+}, \qquad \Omega=\frac{r_-}{r_+},
	\end{align}	
	where $r_+$ and $r_-$ denote the outer and the inner horizon radii respectively. The holomorphic (left moving) and anti-holomorphic (right moving) effective temperatures are given by \cite{Caputa:2013lfa,Banados:1992wn,Banados:1992gq}
	\begin{align}\label{cft_rot_ne_eff_temp}
		T_{\pm}=\frac{1}{\beta_{\pm}}=\frac{r_+ \mp r_-}{2 \pi}.
	\end{align}
	The OEE for this CFT$_2$ may now be obtained by expressing the relevant twist correlator on the twisted cylinder in terms of appropriate correlation function on the complex plane through the conformal map given in \cref{EQ_NE_TRN}.

	For a CFT$_2$ with a conserved charge at a zero temperature, the corresponding bulk dual geometry is described by an extremal rotating BTZ black hole, for which $\beta \to \infty$ and $\Omega \to 1$. In this limit $\beta_+$ goes to infinity and $\beta_-$ remains finite. For this case the twisted cylinder may be mapped to the complex plane through the conformal transformation
	\begin{align}\label{EQ_EX_TR}
		z \rightarrow \omega=z, \qquad \bar{z} \rightarrow \bar{\omega}=\frac{1}{2 r_0} e^{2 r_0 \bar{z}},
	\end{align}
	where $r_0=r_+=r_-$.

	\subsection{OEE for two disjoint intervals}\label{ssec_cft_dis}
	
	We begin with the bipartite mixed state configuration of two disjoint boosted intervals $A=[z_1,z_2]$ and $B=[z_3,z_4]$. The corresponding twist correlator is given by the four-point function in \cref{eq_4pf_dis}, which may be utilized to compute the time dependent OEE for the given bipartite system.

	\subsubsection{Two disjoint intervals at a finite temperature}\label{sn_cft_rot_dis_fin}
	
	 Utilizing \cref{eq_Renyi_oee,eq_vn_oee,eq_4pf,eq_4pf_dis,eq_4pf_dis_cb}, along with the map in \cref{EQ_NE_TRN}, the time dependent OEE for the mixed state of two boosted disjoint intervals at a finite temperature $1/\beta$ in the large $c$ limit may be computed as follows
	 	\begin{align}\label{NE_D_FT}
		S_o(A_1:A_2) &=\frac{c}{6}\log \left[\frac{\beta _+^2 \sinh \left(\frac{\pi  \left(x_{32}+t_{32}\right)}{\beta _+}\right) \sinh \left(\frac{\pi  \left(x_{41}-t_{41}\right)}{\beta _+}\right)}{\pi ^2a^2}\right]+\frac{c}{12} \log \left[\frac{1+\sqrt{\xi_{+}}}{1-\sqrt{\xi_{+}}}\right]\notag\\&+\frac{c}{6}\log \left[\frac{\beta _-^2 \sinh \left(\frac{\pi  \left(x_{32}+t_{32}\right)}{\beta _-}\right) \sinh \left(\frac{\pi  \left(x_{41}-t_{41}\right)}{\beta _-}\right)}{\pi ^2a^2}\right]+\frac{c}{12} \log \left[\frac{1+\sqrt{\xi_{-}}}{1-\sqrt{\xi_{-}}}\right] -\frac{c}{6} \log 2,
	\end{align} 
	where there may be other undetermined OPE coefficients. Here $a$ is the UV cut-off and $\xi_{+}$, $\xi_{-}$ are the cross ratios given by
		\begin{align}
		\xi_{+}=\frac{\sinh \left(\frac{\pi  \left(x_{21}+t_{21}\right)}{\beta _+}\right) \sinh \left(\frac{\pi  \left(x_{43}+t_{43}\right)}{\beta _+}\right)}{\sinh \left(\frac{\pi  \left(x_{31}+t_{31}\right)}{\beta _+}\right) \sinh \left(\frac{\pi  \left(x_{42}+t_{42}\right)}{\beta _+}\right)}\ \ , \qquad   \xi_{-}=\frac{\sinh \left(\frac{\pi  \left(x_{21}-t_{21}\right)}{\beta _-}\right) \sinh \left(\frac{\pi  \left(x_{43}-t_{43}\right)}{\beta _-}\right)}{\sinh \left(\frac{\pi  \left(x_{31}-t_{31}\right)}{\beta _-}\right) \sinh \left(\frac{\pi  \left(x_{42}-t_{42}\right)}{\beta _-}\right)}\ \ .
	\end{align} 
	Note that the OEE computed in \cref{NE_D_FT} may be decomposed into separate left and right moving sectors, characterized by the left and right moving inverse temperatures $\beta_+$ and $\beta_-$ respectively.

	\subsubsection{Two disjoint intervals at zero temperature}
	
	The OEE for two boosted disjoint intervals at zero temperature in the large $c$ limit may be computed in a similar fashion as in \cref{sn_cft_rot_dis_fin} using the map in \cref{EQ_EX_TR} as follows
		\begin{align}\label{EX_DIS}
		S_o(A_1:A_2)&=\frac{c}{6}\log\bigg[\frac{ \left(x_{32}+t_{32}\right)\left(x_{41}+t_{41}\right)}{a^2r_0}\bigg]+\frac{c}{12}\log\bigg[\frac{1+\sqrt{\eta}}{1-\sqrt{\eta}}\bigg]  \notag\\&+\frac{c}{6} \log\bigg[\frac{ \sinh \left(r_0 \left(x_{32}-t_{32}\right)\right)\sinh \left(r_0 \left(x_{41}-t_{41}\right)\right)}{a^2r_0}\bigg]+\frac{c}{12} \log \bigg[\frac{1+\sqrt{\xi_{-}}}{1-\sqrt{\xi_{-}}}\bigg] -\frac{c}{6} \log 2.
	\end{align}
	As in \cref{sn_cft_rot_dis_fin} there may be other undetermined OPE coefficients in \cref{EX_DIS}, and the cross ratios are given by
	\begin{align}
		\eta=\frac{\left(x_{21}+t_{21}\right) \left(x_{43}+t_{43}\right)}{\left(x_{31}+t_{31}\right) \left(x_{42}+t_{42}\right)}\ \ ,\qquad \xi_{-}=\frac{\sinh \left(r_0 \left(x_{21}-t_{21}\right)\right) \sinh \left(r_0 \left(x_{43}-t_{43}\right)\right)}{\sinh \left(r_0 \left(x_{31}-t_{31}\right)\right) \sinh \left(r_0 \left(x_{42}-t_{42}\right)\right)}\ \ .
	\end{align}
	Here $a$ is the UV cut-off. Note that the OEE for this case is characterized by an effective bulk Frolov-Thorne temperature $T_{FT}=r_0/\pi$ \cite{Frolov:1989jh,Caputa:2013lfa}. Furthermore, \cref{EX_DIS} can be decomposed into left and right moving sectors with temperatures zero and $T_{FT}$ respectively.

	\subsection{OEE for two adjacent intervals}
	
	In this subsection we compute the time dependent OEE for two boosted adjacent intervals $A_1=[(x_1,t_1),(x_2,t_2)]$ and $B=[(x_2,t_2),(x_3,t_3)]$, utilizing the three-point function given in \cref{eq_nr_3pf}.

	\subsubsection{Two adjacent intervals at a finite temperature}
	
	Following a prescription similar to \cref{ssec_cft_dis}, the OEE for two boosted adjacent intervals at a finite temperature in the large $c$ limit may be computed as
		\begin{align}\label{NE_ADJ}
		S_o(A_1:A_2)&=\frac{c}{12}\log \left[\frac{\beta _+^3  \sinh \left(\frac{\pi  \left(x_{21}+t_{21}\right)}{\beta _+}\right)\sinh \left(\frac{\pi  \left(x_{32}+t_{32}\right)}{\beta _+}\right)\sinh \left(\frac{\pi  \left(x_{31}+t_{31}\right)}{\beta _+}\right)}{a^3\pi^3 }\right]\notag\\&+\frac{c}{12}\log \left[\frac{\beta _-^3  \sinh \left(\frac{\pi  \left(x_{21}-t_{21}\right)}{\beta _-}\right)\sinh \left(\frac{\pi  \left(x_{32}-t_{32}\right)}{\beta _-}\right)\sinh \left(\frac{\pi  \left(x_{31}-t_{31}\right)}{\beta _-}\right)}{a^3\pi^3 }\right]+\text{constant} ,
	\end{align}
	where $a$ is the UV cut-off and the constant arises from the OPE coefficient in  \cref{eq_nr_3pf}. The separation of the left and right moving sectors is once again evident.

	\subsubsection{Two adjacent intervals at zero temperature}\label{sssec_rot_adj_zt}

The time dependent OEE for two boosted adjacent intervals for a zero temperature CFT$_2$ with a conserved charge in the large central charge approximation may be obtained utilizing the conformal map in \cref{EQ_EX_TR} as follows
	\begin{align}\label{EX_ADJ}
		S_o(A_1:A_2)&=\frac{c}{12}\log\left[ \frac{\left(x_{21}+t_{21}\right)\left(x_{32}+t_{32}\right)\left(x_{31}+t_{31}\right)}{a^3}\right]\notag\\&+\frac{c}{12}\log\left[ \frac{\sinh \left(r_0 \left(x_{21}-t_{21}\right)\right)\sinh \left(r_0 \left(x_{32}-t_{32}\right)\right)\sinh \left(r_0 \left(x_{31}-t_{31}\right)\right)}{a^3r_0^3}\right]+\text{constant},
	\end{align}
	where once again $a$ is the UV cut-off. Note the dependence of the result on the effective bulk Frolov-Thorne temperature $T_{FT}=r_0/\pi$. Once again one may observe from \cref{EX_ADJ} the decomposition into left and right moving sectors with temperatures zero and $T_{FT}$ respectively.

	\subsection{OEE for a single interval}
	
	In this subsection we compute the time dependent OEE for a boosted single interval in CFT$_2$s with a conserved charge at zero and finite temperatures.

	\subsubsection{A single interval at a finite temperature}\label{sssec_cft_ne_sin}
	
In this case it is necessary to consider two boosted auxiliary intervals $B_1=[(-L,t_1),(-l,t_2)]$ and $B_2=[(0,t_3),(L,t_4)]$ on either side of the boosted interval $A=[(-\ell,t_2),(0,t_3)]$, as elaborated in \cref{sssec_nr_cft_sin_ft}. Implementing the bipartite limit $L \to \infty$ gives the desired configuration of a single interval $A$. Utilising \cref{EQ_NE_TRN} and following a procedure similar to \cref{sssec_nr_cft_sin_ft}, the OEE at large $c$ may be obtained as follows
	\begin{align}\label{NE_SIN}
		S_o(A:A^c)&=  \lim_{L \to \infty} \left( \frac{c}{6} \log \left[\frac{\beta _+ \sinh \left(\frac{\pi  \left(2 L+t_{41}\right)}{\beta _+}\right)}{a \pi}\right]+\frac{c}{6} \log \left[\frac{\beta _- \sinh \left(\frac{\pi  \left(2 L-t_{41}\right)}{\beta _-}\right)}{a\pi}\right] \right) \notag\\&
		+\frac{c}{6}\log \left[\frac{\beta _+ \sinh \left(\frac{\pi  \left(\ell+t_{32}\right)}{\beta _+}\right) }{a\pi}\right]-\frac{\pi c \left(\ell+t_{32}\right)}{6\beta _+}+f\left(e^{-\frac{2 \pi  \left(\ell+t_{32}\right)}{\beta _+}}\right)\notag\\&
		+\frac{c}{6}\log \left[\frac{\beta _- \sinh \left(\frac{\pi  \left(\ell-t_{32}\right)}{\beta _-}\right)}{a\pi }\right]-\frac{\pi c \left(\ell-t_{32}\right)}{6\beta _-}+\bar{f}\left(e^{-\frac{2 \pi  \left(\ell-t_{32}\right)}{\beta _-}}\right) \ \ ,
	\end{align}	
	where the terms involving the functions $f,{\bar f}$ are non-universal and depend on the full operator content of the theory which are sub leading in the large $c$ limit and $a$ is the UV cut-off for the CFT$_2$. The decomposition into left and right moving sectors may once again be observed. Specifically, this result incorporates separate thermal entropy terms corresponding to the individual sectors.\footnote{Note that the time contributions to the thermal entropy terms $\pi c(\ell\pm t_{32})/(6\beta_\pm)$ arise possibly due to the rotation of the black hole.}

	\subsubsection{A single interval at zero temperature}
	
Following a procedure similar to \cref{sssec_cft_ne_sin}, for this configuration we consider two boosted auxiliary intervals $B_1$ and $B_2$ on either sides of the interval $A$. Utilising the conformal map given in \cref{EQ_EX_TR}, the OEE at large $c$ upon taking the bipartite limit $L \to \infty$ is obtained as follows
	\begin{align}\label{OEE_EX_SIN}
	S_o(A:A^c)=&\lim_{L \to \infty}\left(\frac{c}{6}\log\left[ \frac{\left(2L+t_{41}\right)}{a}\right]+\frac{c}{6}\log\left[ \frac{\sinh \left(r_0 \left(2L-t_{41}\right)\right)}{ar_0}\right]\right)\notag\\&\frac{c}{6}\log\left[ \frac{\left(\ell+t_{32}\right)}{a}\right]+\frac{c}{6}\log\left[ \frac{\sinh \left(r_0 \left(\ell-t_{32}\right)\right)}{ar_0}\right]\notag\\&-\frac{c \pi }{6}T_{FT}\left(\ell+t_{32}\right)+\bar{f}\left(e^{-2 r_0  \left(\ell-t_{32}\right)}\right) ,
	\end{align}
	where $a$ is the UV cut-off. Once again the presence of an effective Frolov-Thorne temperature $T_{FT}=r_0/\pi$ is observed. Decomposition of \cref{OEE_EX_SIN} into left and right moving sectors having temperatures zero and $T_{FT}$ may also be observed.
	
Note the subtraction of the term $S_{FT}=\frac{c \pi }{6}T_{FT}\left(\ell+t_{32}\right)$ in \cref{OEE_EX_SIN}, which may be interpreted as an effective thermal entropy of the subsystem $A$ associated with the Frolov-Thorne temperature which arises due to the degeneracy of the ground state of the CFT$_2$ with a conserved charge.

	\section{Covariant Holographic OEE in CFT$_2$s with a conserved charge}\label{HOL_OEE_R_SYS}
	
	We now turn our attention to the covariant holographic characterization of the OEE for finite and zero temperature bipartite states in CFT$_2$s with a conserved charge, dual to bulk non-extremal and extremal rotating BTZ black holes. 
	
	For a CFT$_2$ with a conserved charge at a finite temperature, the bulk dual is described by a non-extremal rotating BTZ black hole  given by the metric (with AdS radius set to unity)
	\begin{align}\label{eq_ne_metric}
		\text{d}s^2=-\frac{\left(r^2-r_-^2\right) \left(r^2-r_+^2\right) }{r^2}\text{d}t^2+\frac{r^2 }{\left(r^2-r_-^2\right) \left(r^2-r_+^2\right)}\text{d}r^2+r^2 \left(\text{d}\phi -\frac{r_- r_+ }{r^2}\text{d}t\right)^2 \ .
	\end{align}
	For the metric in \cref{eq_ne_metric}, the coordinate $\phi$ is uncompactified, whereas $r_+$ and $r_-$ describe the outer and inner horizon radii of the black hole respectively. The mass $M$, angular momentum $J$, Hawking temperature $T_H$, and the angular velocity $\Omega$ of the black hole are given by the following relations\footnote{Note that the Hawking temperature $T_H$ corresponds to the temperature $T$ defined in \cref{eq_r_cft_temp}.}
	\begin{align}
		M=r_+^2+r_-^2 , \qquad & \qquad J=2r_+r_-  ,\notag\\ T_H=\frac{1}{\beta}=\frac{r_+^2-r_-^2}{2\pi r_+} ,  \qquad & \qquad \Omega=\frac{r_-}{r_+}  ,
	\end{align}
	whereas the holomorphic and anti-holomorphic effective temperatures are described by [see \cref{cft_rot_ne_eff_temp}]
	\begin{align}\label{bulk_rot_ne_eff_temp}
		T_{\pm}=\frac{1}{\beta_{\pm}}=\frac{r_+ \mp r_-}{2 \pi},
	\end{align}
	where the effective inverse temperatures may also be expressed as $\beta_{\pm}=\beta\left(1\pm i\Omega_E\right)$. The metric in \cref{eq_ne_metric} may be mapped to the flat $\mathbb{R}^{(2,2)}$ through the following embedding relations \cite{Balasubramanian:1998sn,Keski-Vakkuri:1998gmz}
	\begin{align}\label{EM_NE}
		&T_1= \sqrt{\frac{r^2-r_+^2}{r_+^2-r_-^2}} \cosh \left(r_+ x -r_- t\right),\notag\\
		&T_2= \sqrt{\left(\frac{r^2-r_+^2}{r_+^2-r_-^2}-1\right)} \sinh \left(r_+ t-r_- x \right),\notag\\
		&X_1=\sqrt{\frac{r^2-r_+^2}{r_+^2-r_-^2}} \sinh \left(r_+ x -r_- t\right),\notag\\
		&X_2= \sqrt{\left(\frac{r^2-r_+^2}{r_+^2-r_-^2}-1\right)} \cosh \left(r_+ t-r_- x\right) .
	\end{align}

	On the other hand for a zero temperature CFT$_2$ with a conserved charge, the bulk dual is characterized by an extremal rotating BTZ black hole described by the metric
	\begin{align}\label{eq_e_metric}
		\text{d}s^2=-\frac{\left(r^2-r_0^2\right)^2 }{r^2}\text{d}t^2+\frac{r^2 }{\left(r^2-r_0^2\right)^2}\text{d}r^2+r^2 \left(\text{d}\phi -\frac{r_0^2 }{r^2}\text{d}t\right)^2  .
	\end{align}
	Note that the extremal limit is represented by $r_+=r_-=r_0$ such that the mass is equal to the angular momentum $(M=J=2r_0^2)$ and the Hawking temperature is zero $(T_H=0)$. The metric in \cref{eq_e_metric} may be mapped to the Poincar\'e coordinates using the following embedding relations \cite{Keski-Vakkuri:1998gmz}
	\begin{align}\label{EM_EX}
		&T= \frac{1}{2} \left[-\frac{r_0}{r^2-r_0^2}-\frac{e^{2 r_0 (x-t)}}{2 r_0}+t+x\right],\notag\\
		&X= \frac{1}{2} \left[-\frac{r_0}{r^2-r_0^2}+\frac{e^{2 r_0 (x-t)}}{2 r_0}+t+x\right],\notag\\
		&z= \frac{e^{r_0 (x-t)}}{\sqrt{r^2-r_0^2}}  .
	\end{align}
	Note that for both non-extremal and extremal rotating BTZ black holes, the conformal boundary is located at $r=r_c$, where $r_c$ is the bulk IR cut-off related to the boundary UV cut-off $a$ as $r_c \sim 1/a$.

	\subsection{Holographic OEE for two disjoint intervals}
	
	In this subsection we compute the covariant holographic OEE for the configuration of two boosted disjoint intervals in CFT$_2$s dual to non-extremal and extremal rotating BTZ black holes.
	
	\subsubsection{Two disjoint intervals for a non-extremal BTZ black hole}

We begin with two boosted disjoint intervals $A_1=[(x_1,t_1),(x_2,t_2)]$ and $A_2=[(x_3,t_3),(x_4,t_4)]$ in a thermal CFT$_2$ dual to a non-extremal rotating BTZ black hole. The EWCS for this configuration may be computed utilizing the embedding relations in \cref{EM_NE} and the prescription in \cref{eq_nr_dis_ew} as follows	
	\begin{align}\label{EW_NE_DIS}
		E_W(A_1:A_2)=	\frac{c}{12} \log \left[\frac{1+\sqrt{\xi_{+}}}{1-\sqrt{\xi_{+}}}\;\right]+\frac{c}{12} \log \left[\frac{1+\sqrt{\xi_{-}}}{1-\sqrt{\xi_{-}}}\; \right] \ \ .
	\end{align}
	Here  $\eta_{ne}$ and $\bar{\eta}_{ne}$ are the cross ratios given as
	\begin{align}
		\xi_{+}=\frac{\sinh \left(\frac{\pi  \left(x_{21}+t_{21}\right)}{\beta _+}\right) \sinh \left(\frac{\pi  \left(x_{43}+t_{43}\right)}{\beta _+}\right)}{\sinh \left(\frac{\pi  \left(x_{31}+t_{31}\right)}{\beta _+}\right) \sinh \left(\frac{\pi  \left(x_{42}+t_{42}\right)}{\beta _+}\right)} ,\quad 
		\xi_{-}=\frac{\sinh \left(\frac{\pi  \left(x_{21}-t_{21}\right)}{\beta _-}\right) \sinh \left(\frac{\pi  \left(x_{43}-t_{43}\right)}{\beta _-}\right)}{\sinh \left(\frac{\pi  \left(x_{31}-t_{31}\right)}{\beta _-}\right) \sinh \left(\frac{\pi  \left(x_{42}-t_{42}\right)}{\beta _-}\right)} \ \ .
	\end{align}
	The holographic EE may be computed using \cref{eq_holo_ee} as
		\begin{align}\label{EE_NE_DIS}
		S(A_1\cup A_2)&=\frac{c}{6} \log \left[\frac{\beta _+^2 \sinh \left(\frac{\pi  \left(x_{41}+t_{41}\right)}{\beta _+}\right)\sinh \left(\frac{\pi  \left(x_{32}+t_{32}\right)}{\beta _+}\right) }{\pi ^2 a^2}\right]\notag\\&+\frac{c}{6} \log \left[\frac{\beta_-^2 \sinh \left(\frac{\pi  \left(x_{41}-t_{41}\right)}{\beta _-}\right)\sinh \left(\frac{\pi  \left(x_{32}-t_{32}\right)}{\beta _-}\right) }{\pi ^2 a^2}\right] .
	\end{align}
	The holographic OEE may now be obtained from the relation in \cref{eq_holodual} using \cref{EW_NE_DIS,EE_NE_DIS}. Interestingly the holographic results match with the field theory replica technique results described in \cref{NE_D_FT} at large $c$ up to certain constant terms which once again verifies the holographic duality. Also note that the resulting holographic OEE may be decomposed into left and right moving sectors with inverse temperature $\beta_+$ and $\beta_-$ respectively.

	\subsubsection{Two disjoint intervals for an extremal BTZ black hole}

For the configuration of two boosted disjoint intervals in a zero temperature CFT$_2$ dual to an extremal rotating BTZ black hole, the EWCS may be obtained utilizing the embedding relations in \cref{EM_EX} and the prescription in \cref{eq_nr_dis_ew} as follows
	\begin{align}\label{EW_EX_DIS}
		E_W(A_1:A_2)=\frac{c}{12} \log \left[\frac{1+\sqrt{\eta}}{1-\sqrt{\eta}}\right]+\frac{c}{12} \log \left[\frac{1+\sqrt{\xi_{-}}}{1-\sqrt{\xi_{-}}}\right] ,
	\end{align}
	where $\eta$ and $\xi_{-}$ are cross ratios given by
	\begin{align*}
		\eta=\frac{\sinh \left(r_0 \left(x_{21}-t_{21}\right)\right) \sinh \left(r_0 \left(x_{43}-t_{43}\right)\right)}{\sinh \left(r_0 \left(x_{31}-t_{31}\right)\right) \sinh \left(r_0 \left(x_{42}-t_{42}\right)\right)}\;, \qquad \xi_{-}=\frac{\left(x_{21}+t_{21}\right) \left(x_{34}-t_{34}\right)}{\left(x_{31}-t_{31}\right) \left(x_{42}-t_{42}\right)} \;.
	\end{align*}
	The relevant holographic entanglement entropy may be computed from \cref{eq_holo_ee} as
	\begin{align}\label{EE_NEX_DIS}
		S(A_1\cup A_2)&=\frac{c}{6}\log \left[\frac{\left(x_{32}+t_{32}\right)\left(x_{41}+t_{41}\right)}{a^2}\right]\notag\\&+\frac{c}{6}\log \left[\frac{\sinh \left(r_0 \left(x_{32}-t_{32}\right)\right)\sinh \left(r_0 \left(x_{41}-t_{41}\right)\right)}{a^2 r_0^2}\right].
	\end{align}
	The holographic OEE  may now be obtained from \cref{eq_holodual,EW_EX_DIS,EE_NEX_DIS}, which is dependent on the effective Frolov-Thorne temperature $T_{FT}=r_0/\pi$. It is observed that our holographic result matches with the corresponding field theory replica technique result obtained in the large $c$ limit in \cref{EX_DIS} modulo certain constants substantiating the consistency of our holographic construction. Furthermore note that once again the holographic OEE may be decomposed into left and right moving sectors with temperature zero and $T_{FT}$ respectively.

	\subsection{Holographic OEE for two adjacent intervals}
	
	In this subsection we compute the holographic OEE for two boosted adjacent intervals $A_1=[(x_1,t_1),(x_2,t_2)]$ and $A_2=[(x_2,t_2),(x_3,t_3)]$ in CFT$_2$s dual to non-extremal and extremal rotating BTZ black holes.

	\subsubsection{Two adjacent intervals for a non-extremal BTZ black hole}
	
	Using \cref{eq_nr_adj_ew} and the embedding relations in \cref{EM_NE}, it is possible to compute the EWCS for two boosted adjacent intervals in a CFT$_2$ with a conserved charge at a finite temperature dual to a bulk non-extremal rotating BTZ black hole as
	\begin{align}\label{EW_NE_ADJ}
		E_W(A_1:A_2)&=\frac{c}{12}\log \left[\frac{\beta _+ \sinh \left(\frac{\pi  \left(x_{21}+t_{21}\right)}{\beta _+}\right)\sinh \left(\frac{\pi  \left(x_{32}+t_{32}\right)}{\beta _+}\right)}{a \pi \sinh \left(\frac{\pi  \left(x_{31}+t_{31}\right)}{\beta _+}\right)}\right]\notag\\&
		+\frac{c}{12}\log \left[\frac{\beta _- \sinh \left(\frac{\pi  \left(x_{21}-t_{21}\right)}{\beta _-}\right)\sinh \left(\frac{\pi  \left(x_{32}-t_{32}\right)}{\beta _-}\right)}{a \pi \sinh \left(\frac{\pi  \left(x_{31}-t_{31}\right)}{\beta _-}\right)}\right]+\frac{c}{6}\log 2 .
	\end{align}
	Utilizing \cref{eq_holo_ee}, the holographic EE may be given as
	\begin{align}\label{EE_ADJ_NEX}
		\hspace{-1.5cm}S(A_1\cup A_2)=\frac{c}{6}\log\left[\frac{\beta _+\sinh \left(\frac{\pi  \left(x_{31}+t_{31}\right)}{\beta _+}\right)}{a\pi}\right]+\frac{c}{6}\log\left[\frac{\beta _- \sinh \left(\frac{\pi  \left(x_{31}-t_{31}\right)}{\beta _-}\right)}{a\pi}\right]  .
	\end{align}
Once again it is observed that the holographic OEE computed through \cref{eq_holodual,EW_NE_ADJ,EE_ADJ_NEX} for this configuration matches with the large $c$ limit field theory results obtained in \cref{NE_ADJ} modulo certain constants, which serves to substantiate our construction. Note again that the holographic OEE may be decomposed into the left and right moving sectors with inverse temperatures $\beta_+$ and $\beta_-$ respectively.

	\subsubsection{Two adjacent intervals for an extremal BTZ black hole}
	
	For the configuration of two boosted adjacent intervals in a CFT$_2$ at zero temperature dual to an extremal rotating BTZ black hole, the EWCS between $A_1$ and $A_2$, and $S(A_1\cup A_2)$ may be computed as follows
	\begin{align}\label{EW_EX_ADJ}
		E_W(A_1:A_2)&=\frac{c}{12}\log\left[ \frac{\left(x_{21}+t_{21}\right)\left(x_{32}+t_{32}\right)}{a \left(x_{31}+t_{31}\right)}\right]\notag\\&
		+\frac{c}{12}\log\left[ \frac{\sinh \left(r_0 \left(x_{21}-t_{21}\right)\right)\sinh \left(r_0 \left(x_{32}-t_{32}\right)\right)}{ar_0 \sinh \left(r_0 \left(x_{31}-t_{31}\right)\right)}\right]+\frac{c}{6} \log 2  ,
	\end{align}
	\begin{align}\label{EE_ADJ_EX}
		S(A_1\cup A_2)=\frac{c}{6}\log\left[ \frac{\left(x_{31}+t_{31}\right)}{a}\right]+\frac{c}{6}\log\left[ \frac{\sinh \left(r_0 \left(x_{31}-t_{31}\right)\right)}{ar_0}\right] .
	\end{align}
	Once again, the holographic OEE obtained by combining \cref{EW_EX_ADJ,EE_ADJ_EX,eq_holodual} for this case matches with the field theory results at large $c$ described in \cref{EX_ADJ}, up to certain constant terms. Furthermore, note that similar to the field theory results in \cref{sssec_rot_adj_zt}, the holographic OEE is observed to depend on the effective bulk Frolov-Thorne temperature $T_{FT}=r_0/\pi$ and may also be decomposed into left and right moving sectors with temperatures zero and $T_{FT}$ respectively.

	\subsection{Holographic OEE for a single interval}
	
	Finally in this subsection we consider a boosted single interval $A$ in CFT$_2$s dual to non-extremal and extremal rotating BTZ black holes. As discussed in \cref{ssec_nr_holo_sin}, the procedures for computing the EWCS for the non-extremal and extremal cases are different.

	\subsubsection{A single interval for a non-extremal BTZ black hole}\label{sssec_hol_ne_sing}
	
For this configuration it is required to consider two boosted auxiliary intervals $B_1=[(-L,t_1),(-l,t_2)]$ and $B_2=[(0,t_3),(L,t_4)]$ sandwiching the interval $A=[(-\ell,t_2),(0,t_3)]$ under consideration. Recall that implementing the bipartite limit $L \to \infty$ gives the configuration of a single interval $A$.

The upper bound to the EWCS as described in \cref{eq_nr_sin_ew_n} may then be computed as	
	\begin{align}\label{EW_NE_SIN}
		\tilde{E}_W(A:A^c) & =\frac{c}{6}\log \left[\frac{\beta _+ \sinh \left(\frac{\pi  \left(\ell+t_{32}\right)}{\beta _+}\right) }{a\pi}\right]-\frac{\pi c \left(\ell+t_{32}\right)}{6\beta _+}\notag\\&
		+\frac{c}{6}\log \left[\frac{\beta _- \sinh \left(\frac{\pi  \left(\ell-t_{32}\right)}{\beta _-}\right)}{a\pi }\right]-\frac{\pi c \left(\ell-t_{32}\right)}{6\beta _-}+\frac{c}{6}\log 4,
	\end{align}
	where the bipartite limit $L \to \infty$ has already been implemented. The holographic EE is given as
	\begin{align}\label{HEE_NE_SIN}
S(A\cup A^c) = \lim_{L \to \infty} \left( \frac{c}{6} \log \left[\frac{\beta _+ \sinh \left(\frac{\pi  \left(2 L+t_{41}\right)}{\beta _+}\right)}{a\pi}\right]+\frac{c}{6} \log \left[\frac{\beta _- \sinh \left(\frac{\pi  \left(2 L-t_{41}\right)}{\beta _-}\right)}{a\pi}\right] \right).
	\end{align}
	As earlier we note that the holographic OEE computed using \cref{eq_holodual,EW_NE_SIN,HEE_NE_SIN} matches with the corresponding large $c$ field theory result described in \cref{NE_SIN} up to a constant term and  may be decomposed into left and right moving sectors with inverse temperatures $\beta_+$ and $\beta_-$ respectively.

	\subsubsection{A single interval for an extremal BTZ black hole}
	
Following the procedure as elaborated in \cref{sssec_hol_ne_sing}, for this case we consider auxiliary intervals on either side of the interval $A$. Using the embedding relations given is \cref{EM_EX}, the upper bound to the EWCS may be computed as 
	\begin{align}\label{eq_ew_e_sin}
	\tilde{E}_W(A:A^c)=	
	\frac{c}{6}\log\left[ \frac{\left(\ell+t_{32}\right)}{a}\right]+\frac{c}{6}\log\left[ \frac{\sinh \left(r_0 \left(\ell-t_{32}\right)\right)}{ar_0}\right]-\frac{c \pi }{6}T_{FT}\left(\ell+t_{32}\right)+	\frac{c}{6}\log4,
		\end{align}
where we have already implemented the bipartite limit $L \to \infty$. The holographic EE is given as 
\begin{align}\label{eq_hee_e_sin}
	S(A \cup A^c)=\lim_{L \to \infty}\left(\frac{c}{6}\log\left[ \frac{\left(2L+t_{41}\right)}{a}\right]+\frac{c}{6}\log\left[ \frac{\sinh \left(r_0 \left(2L-t_{41}\right)\right)}{ar_0}\right]\right).
\end{align}
The holographic OEE may be computed using \cref{eq_holodual,eq_ew_e_sin,eq_hee_e_sin}, which matches with the corresponding field theory results up to a constant. Also note the dependence of the holographic OEE on the effective Frolov-Thorne temperature $T_{FT}=r_0/\pi$. From the holographic perspective, the effective thermal entropy term in \cref{eq_ew_e_sin} associated with $T_{FT}$ indicates an emergent thermodynamic behavior of the rotating extremal BTZ black hole at zero temperature due to the non-zero entropy of degeneracy. The holographic OEE in this case may also be decomposed into left and right moving sectors with temperature zero and $T_{FT}$ respectively.

\section{Summary and conclusions}\label{SUMMaRY}

To summarize, in this article we have advanced a covariant holographic construction for the OEE of time dependent bipartite states in CFT$_2$s dual to bulk AdS$_3$ geometries. In this context a replica technique has been developed to obtain the OEE for time dependent bipartite states in such CFT$_2$s. This technique was utilized to compute the OEE for bipartite state configurations described by  two boosted disjoint intervals, two boosted adjacent intervals, and a boosted single interval, in zero and finite temperature and finite size CFT$_2$s dual to bulk Poincar\'e, global AdS$_3$ and BTZ black hole geometries respectively. Subsequently we have obtained the OEE for the above bipartite states through a holographic duality described in the literature, involving the corresponding EWCS computed for the bulk dual geometries. Interestingly our field theory replica technique results match with the holographic computations in the large central charge limit modulo certain constants arising from OPE coefficients and/or conformal block expansions substantiating the holographic duality in the literature.

Subsequently we have extended the above analysis for the OEE of time dependent bipartite states to zero and finite temperature holographic CFT$_2$s with a conserved charge dual to bulk stationary geometries described by extremal and non-extremal rotating BTZ black holes. In this connection we have also obtained the OEE for two boosted disjoint intervals, two boosted adjacent intervals, and a boosted single interval, in the dual CFT$_2$s through a replica technique. Furthermore we have computed the corresponding covariant OEE through the holographic duality involving the EWCS for the dual bulk geometries mentioned above. Once again we were able to demonstrate the agreement of the holographic computations with the replica technique results in the large central charge limit modulo certain constants in a strong substantiation of the duality mentioned earlier.

Interestingly our results for the covariant OEE of a single interval in a finite temperature CFT$_2$ involved the elimination of the relevant thermal entropy term in conformity with its definition as a mixed state correlation measure in quantum information theory. For the other mixed state configurations involving  two disjoint and adjacent intervals this elimination is more subtle involving the mutual cancellation of the corresponding thermal entropy terms in the expression for the OEE.

Our covariant holographic construction for the OEE naturally constitutes a significant technique for the investigation of the mixed state entanglement structure of time dependent systems in diverse areas such as many body theory and the black hole information loss paradox in the quantum extremal island framework. Although in principle our covariant holographic proposal for the OEE may be extended to generic AdS$_{d+1}$/CFT$_d$ scenarios, the technical difficulties with the construction of the relevant (HRT) surfaces and the corresponding extremal EWCS in this context are well known. Hence it is a challenging open issue to find substantial examples for such a higher dimensional extension of our covariant construction. Furthermore a proof for the holographic duality between the OEE and the extremal EWCS based on a bulk gravitational path integral and its covariant extension is also another difficult but significant question which requires investigation. We hope to return to some of these issues in the near future.

\section*{Acknowledgments}

We would like to thank Debarshi Basu, Vinayak Raj and Himanshu Chourasiya for useful discussions. The work of Gautam Sengupta is supported in part by the Dr. Jagmohan Garg Chair Professor position at the Indian Institute of Technology, Kanpur. The work of Saikat Biswas is supported by the Council of Scientific and Industrial Research (CSIR) of India under Grant No. 09/0092(12686)/2021-EMR-I.

\section*{Data availability statement}

No Data associated in the manuscript.

\section*{Conflict of interest}
The authors declare that there are no potential financial or non-financial conflict of interests.
\bibliographystyle{utphys}
\bibliography{reference}
\end{document}